\documentclass[
reprint,
superscriptaddress,
 amsmath,
 amssymb,
 aps,
prl,
longbibliography,
]{revtex4-1}

\usepackage[normalem]{ulem}

\usepackage{graphicx}
\usepackage{dcolumn}
\usepackage{bm}

\usepackage{color}
\usepackage{colordvi}  
\definecolor{Red}{rgb}{1,0,0}

\newcommand{\bra}[1]{\left\langle #1 \right|}
\newcommand{\ket}[1]{\left| #1 \right\rangle}

\makeatletter
\def\authornote{\xdef\@thefnmark{$\dagger$}\@footnotetext}
\makeatother

\begin{document}

\title{Quantum dot single photon sources with ultra-low multi-photon probability}

\author{Lukas Hanschke$^\dagger$}
\affiliation{Walter Schottky Institut and Physik Department, Technische Universit\"at M\"unchen, 85748 Garching, Germany}
\author{Kevin A. Fischer$^\dagger$}
\affiliation{E. L. Ginzton Laboratory, Stanford University, Stanford, California 94305, USA}
\authornote{These authors contributed equally.}
\author{Stefan Appel}
\affiliation{Walter Schottky Institut and Physik Department, Technische Universit\"at M\"unchen, 85748 Garching, Germany}
\author{Daniil Lukin}
\affiliation{E. L. Ginzton Laboratory, Stanford University, Stanford, California 94305, USA}
\author{Jakob Wierzbowski}
\affiliation{Walter Schottky Institut and Physik Department, Technische Universit\"at M\"unchen, 85748 Garching, Germany}
\author{Shuo Sun}
\affiliation{E. L. Ginzton Laboratory, Stanford University, Stanford, California 94305, USA}
\author{Rahul Trivedi}
\affiliation{E. L. Ginzton Laboratory, Stanford University, Stanford, California 94305, USA}
\author{Jelena Vu\v{c}kovi\'c}
\affiliation{E. L. Ginzton Laboratory, Stanford University, Stanford, California 94305, USA}
\author{Jonathan J. Finley}
\affiliation{Walter Schottky Institut and Physik Department, Technische Universit\"at M\"unchen, 85748 Garching, Germany}
\author{Kai M\"uller}
\email{kai.mueller@wsi.tum.de}
\affiliation{Walter Schottky Institut and Physik Department, Technische Universit\"at M\"unchen, 85748 Garching, Germany}

\date{\today}

\begin{abstract}
High-quality sources of single photons are of paramount importance for quantum communication, sensing and metrology. To these ends, resonantly excited two-level systems based on self-assembled quantum dots have recently generated widespread interest. Nevertheless, we have recently shown that for resonantly excited two-level systems, emission of a photon during the presence of the excitation laser pulse and subsequent re-excitation results in a degradation of the obtainable single-photon purity. Here, we demonstrate that generating single photons from self-assembled quantum dots with a scheme based on two-photon excitation of  the biexciton strongly suppresses the re-excitation. Specifically, the pulse-length dependence of the multi-photon error rate reveals a quadratic dependence in contrast to the linear dependence of resonantly excited two-level systems, improving the obtainable multi-photon error rate by several orders of magnitude for short pulses. We support our experiments with a new theoretical framework and simulation methodology to understand few-photon sources.
\end{abstract}

\pacs{Valid PACS appear here}

\maketitle

\section{Introduction}

Two-level systems (2LS) provided by excitonic transitions in self-assembled quantum dots (QDs) are commonly used on-demand sources for high-quality single photons \cite{michler, Senellart2017}. Crucially, resonant excitation enables nearly transform-limited linewidth \cite{Prechtel2013, Hansom2014, Kuhlmann2015} and high photon indistinguishability \cite{He2013}. Combined with nanoresonators, single-photon sources with high emission rates and collection efficiency have been demonstrated \cite{Somaschi2016, Unsleber2016, Ding2016, He2016, Mueller2016, Wang2016, Loredo2016, Iles-Smith2017} and are now being incorporated into quantum information processors. For example, a solitary high-quality QD source was recently used in exciting demonstrations to create a train of single photons, which were temporally multiplexed to the input of a Boson Sampler \cite{Loredo2017, Wang2017}. Boson sampling with this source, to date, has provided one of the best experimental validations of optical quantum computing. The quality of the experimental data in this, and future optical quantum information processors, ultimately relies on the ability of the source to emit precisely one photon when triggered by a laser pulse. However, it has recently been shown (from our work \cite{Fischer2016, Fischer2017, Fischer2017-2} and others \cite{dada2016indistinguishable}) that resonant excitation of a 2LS provides a fundamental limitation to the error rate of the single-photon source and hence the information processor. This results from the emission of a photon during the presence of the excitation pulse which leads to re-excitation and multi-photon emission.

In this letter, we investigate an alternative scheme which is based on a four-level system, given by the biexciton-exciton ladder in a QD, and demonstrate that it facilitates significantly higher single-photon purity than a resonantly driven 2LS due to dramatically reduced re-excitation. At the same time, it maintains a simple implementation and high single-photon generation rates. Moreover, it enables an even higher brightness since it eliminates the need for polarization suppression of the excitation laser.

\section{Results}

We first provide detailed experimental results supporting that the bi-excitonic system is a superior single-photon source over a two-level system due to re-excitation. Second, we provide a new theoretical analysis for photon sources, using the biexcitonic system as an example.

\subsection{Experimental results}

\begin{figure}[!t]
  \includegraphics[width=\columnwidth]{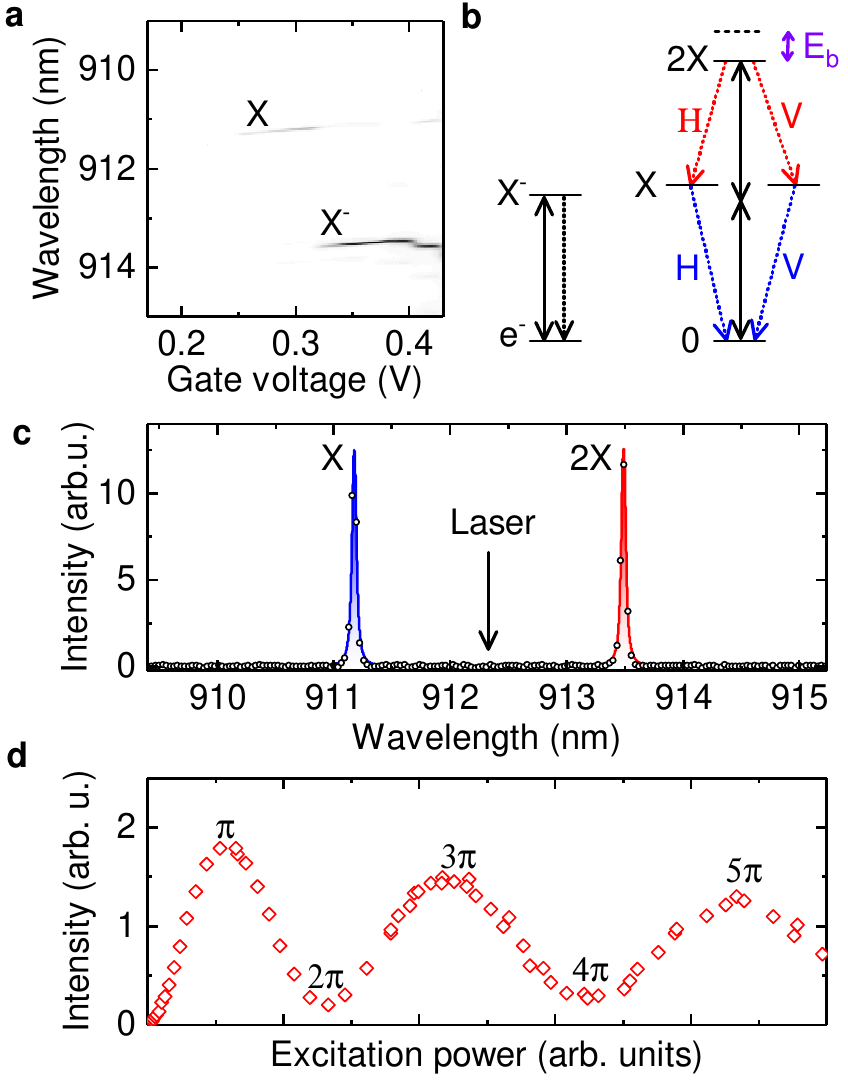}
  \caption{\textbf{Two-photon excitation scheme.} (a) Gate voltage dependent photoluminescence. (b) Schematic illustration of resonant excitation of a 2LS and two-photon excitation of 2X. Solid arrows represent laser drive, while dotted arrows represent spontaneous emission. (c) Example spectrum for two-photon excitation of 2X at 912.34 nm. (d) Rabi oscillations between $\ket{0}$ and $\ket{2X}$, measured from 2X luminescence.}
  \label{figure:1}
\end{figure}

The sample consists of InGaAs QDs of low areal density ($<1\mu m^{-2}$) embedded in the intrinsic region of an n-i Schottky diode. The QDs are grown at a distance of 35nm from the n-doped region which allows control of the charge occupancy of the QDs. A field-dependent photoluminescence measurement is presented in Fig. \ref{figure:1}a and confirms clear charge stability plateaus for the neutral exciton transitions $X$ and emission from a negatively charged trion ($X^-$). The latter can be used as a true two-level system with emission rate $\gamma_{X^-}$ (Fig. \ref{figure:1}b---left) while the former is part of a four-level system given by the biexciton $2X$, exciton $X$ and ground state $0$ (Fig. \ref{figure:1}b---right). Due to anisotropy in QD shape, the exchange interaction results in two $X$ levels where one couples $2X$ and $0$ with horizontal polarization and the other with vertical polarization \cite{Bayer2002, Finley2002}. Depending on the specific type of QD, the two $X$ levels are non-degenerate with a fine structure splitting of $0-100\;\mu\text{eV}$. This system is well-known for the generation of entangled photon pairs \cite{Akopian2006, Mueller2014, Huber2017}. Due to the Coulomb interaction, the energy of $2X$ is detuned from twice the $X$ energy by the binding energy $E_{b}$. Therefore, $2X$ can be excited via a two-photon process where the laser energy is detuned from $X$  by $E_{b} /2$ \cite{Mueller2014, Ardelt2014}. The emission rates of this system are $\gamma_{2X}$ and $\gamma_{X}$. A typical spectrum for two-photon excitation of $2X$ is presented in Fig. \ref{figure:1}c and confirms identical intensities for $2X$ and $X$ emission as expected. The dependence of the emission intensity on the excitation power is presented in Fig. \ref{figure:1}d for exciting with 3 ps long pulses and reveals clean Rabi oscillations. Time-resolved measurements reveal lifetimes of 260 ps for the emission from $X$ and 173 ps for the emission from $2X$.

\begin{figure}[!t]
  \includegraphics[width=\columnwidth]{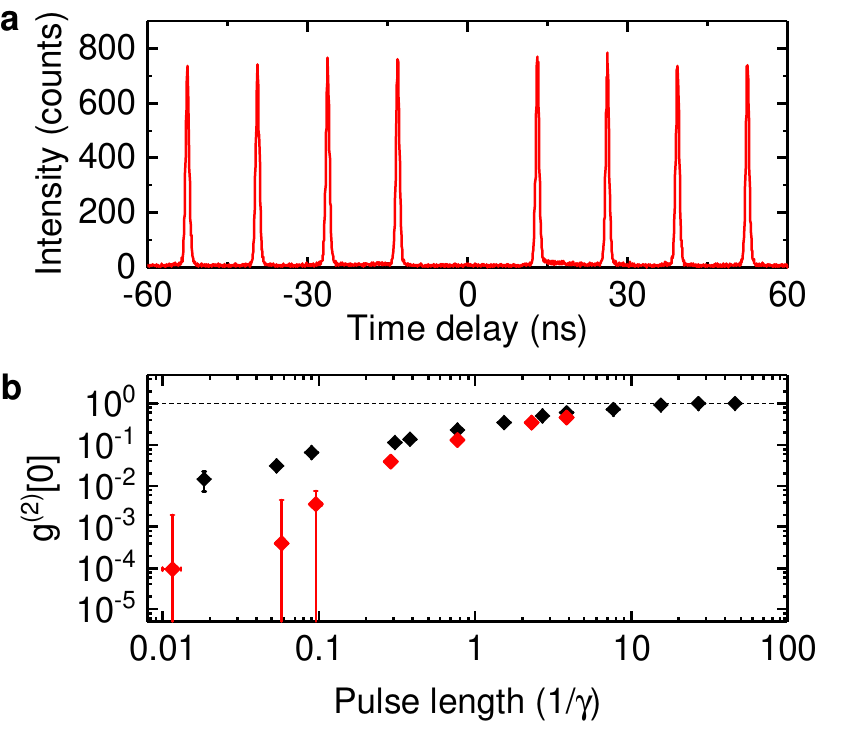}
  \caption{\textbf{Measured degree of second-order coherence.} (a) Example of a measurement for two-photon excitation and filtering on the 2X emission using a pulse length of 3 ps. (b) Measured values of $g^{(2)}[0]$ as a function of the pulse length for a resonantly driven two-level system (black) and two-photon excitation of 2X (red). Dashed line represents Poissonian statistics of driving laser.}
  \label{figure:2}
\end{figure}

To generate single photons from this system, the emission has simply to be frequency filtered to the $2X$ or $X$ transition. The single-photon purity is quantified by the measured degree of second-order coherence $g^{(2)}[0]=\langle n(n-1)\rangle/\langle n\rangle^2$ where $n$ is the number of photons per pulse. We performed measurements of $g^{(2)}[0]$ for two-photon excitation and frequency filtered detection on $2X$ using a standard HBT setup. An example measurement for $3\,\text{ps}$ long excitation pulses of area $\pi$ is presented in Fig. \ref{figure:2}a. The obtained values for $g^{(2)}[0]$ are so low that fitting the data with a series of peaks does not yield a value. Therefore, we integrate the data over an interval that contains the complete peaks (2.6 ns) and compare the integrated counts around zero time delay to the average intensity of the peaks away from zero delay. After subtracting a constant dark count background we obtain a value of $g^{(2)}[0] = 9.4\cdot10^{-5} \pm 1.9 \cdot10^{-3}$.

The measured value of $g^{(2)}[0]$ is lower than values obtained for a resonantly driven 2LS \cite{Fischer2016}. While the upper limit is dominated by the error which results from the dark counts of the avalanche photodiodes used here, very recently Schweickert et al have reported a value of $g^{(2)}[0] = (7.5\pm1.6)\cdot10^{-5}$ using the same scheme but superconducting detectors with negligible dark count rates  \cite{Schweickert2017}. As discussed above, for the resonantly excited 2LS, $g^{(2)}[0]$ is limited by re-excitation that is enabled by emission of a photon during the presence of the pulse. In contrast, for a two-photon excitation of $2X$, re-excitation is strongly suppressed. Because the laser is far detuned from the $2X$ transition by $E_{b} /2$ re-excitation following the excitation of $2X$ can only occur after the cascade $\ket{2X}\rightarrow \ket{X}\rightarrow \ket{0}$ has returned the system to the ground state.

Since the re-excitation probability depends on the pulse length $T$, we performed measurements of $g^{(2)}[0]$ for different values of $T$. The results are presented as red datapoints in Fig. \ref{figure:2}b. For comparison, the values obtained for a resonantly driven 2LS formed by the $X^-$ transition of the same QD are presented in Fig. \ref{figure:2}b as black datapoints (reproduced from Ref. \cite{Fischer2017-2}). Thereby, the pulse lengths are normalized to $\gamma_{X^-}$ and $\gamma_{X}$, respectively. Note, for the 2LS data and very short pulses, $g^{(2)}[0]$ was corrected for an imperfect suppression of the excitation laser which can be quantified by electrically detuning the $X^-$ transition \cite{Fischer2017-2}. For both cases  $g^{(2)}[0]$ increases  with pulse length and asymptotically approaches the classical limit of 1 for long pulses. Crucially, for all measured pulse lengths, the values obtained from the two-photon excitation scheme are significantly lower than the resonantly excited 2LS. The  improvement in $g^{(2)}[0]$ amounts to an improvement of several orders of magnitude for sufficiently short pulses. A power law fit in the short pulse regime (not shown) results coefficients of $0.73\pm0.07$ for the resonantly driven 2LS and $1.89\pm0.40$ for the two-photon excitation scheme. This indicates a scaling behavior of approximately $g^{(2)}[0] \propto T  \gamma$  for the resonantly driven 2LS and $g^{(2)}[0] \propto (T  \gamma)^2$ for the two-photon excitation scheme, which will be confirmed in theoretical considerations below. Slight deviations of the experimental data from this behavior can be attributed to changes in the pulse shape when increasing the pulse length.

\subsection{Theoretical results}

Next, we gain insight into the behavior through a theoretical study of the emission from an ideal 2LS and 2X system. First, consider an ideal two-level system \cite{Shore2011-fz,Fischer2017-2}, with a ground state $\ket{e^-}$ and an excited state $\ket{X^-}$. Suppose the system is driven by an optical pulse starting at $t=0$, resonant with the $\ket{e^-}\leftrightarrow\ket{X^-}$ transition and where the rotating wave approximation holds. As a function of the interacted pulse area
\begin{equation}
A_\text{2LS}(t) = {\int_{0}^t \mathop{\textrm{d} t'} \mu\cdot E(t')/\hbar},\label{eq:area}
\end{equation}
where $E(t')$ is the envelope of the pulse's electric field and $\mu$ the system's electric dipole moment, the system undergoes coherent oscillations between its ground $\ket{e^-}$ and excited $\ket{X^-}$ states. If the system is initially prepared in the ground state, as is typical in cryogenic experiments, the probability of the system being in the excited state $\textrm{P}_{X^-}(A(t))$ shows Rabi oscillations that are nearly sinusoidal
\begin{equation}
\textrm{P}_{X^-}(A(t))\approx \textrm{sin}^2(A(t)/2),
\end{equation}
for excitation by a short pulse relative to the spontaneous emission time of the 2LS. The Rabi oscillations are captured by the Hamiltonian (in a reference frame rotating at the laser frequency)
\begin{equation}
H_\text{2LS}(t) =  \frac{\mu\cdot E(t)}{2}\left(\ket{e^-}\bra{X^-}+\ket{X^-}\bra{e^-}\right),
\end{equation}
where $\sigma=\ket{e^-}\bra{X^-}$ is the system's dipole operator.

Second, to model the 2X system we will actually use only a three-level system (3LS) with levels labeled as $\ket{0}$, $\ket{X'}$ and $\ket{2X}$. Although there is strictly no transformation that makes these systems equivalent, if the polarization of photons emitted is disregarded in the photon counting procedure, then the behavior of the 3LS mirrors that of the 2X system. Since a two-photon transition excites the system $\ket{0}\leftrightarrow\ket{2X}$ via the intermediate state $\ket{X'}$, the system undergoes Rabi oscillations that scale linearly with the pulse power rather than the field. Hence,
\begin{equation}
A_\text{3LS}(t) = {\int_{0}^t \mathop{\textrm{d} t'} \frac{ (\mu\cdot E(t'))^2 }{\hbar E_b}}
\end{equation}
and
\begin{equation}
H_\text{3LS}(t) =  \frac{\left(\mu\cdot E(t)\right)^2}{2E_b}\left(\ket{0}\bra{2X}+\ket{2X}\bra{0}\right)
\end{equation}
where $\sigma\in\{\ket{0}\bra{X'},\ket{X'}\bra{2X}\}$ are the system's dipole operators. The operator $\ket{0}\bra{2X}$ only appears after adiabatic elimination of the intermediate state.

The dynamics of the systems under spontaneous emission into Markovian reservoirs are captured in the density operator for the systems, whose evolutions can be written in terms of a Liouvillian as
\begin{eqnarray}
\rho(t_1)&=&\mathcal{V}(t_1,t_0)\rho(t_0)\nonumber\\
&=& T_\leftarrow\text{exp}\left[\int_{t_0}^{t_1}\mathop{\text{d}t}\mathcal{L}(t)\right]\rho(t_0),\label{eq:Liouvillian}
\end{eqnarray}
where $T_\leftarrow$ is the chronological operator which orders the infinitesimal products in Eq. \ref{eq:Liouvillian}. The Liouvillian is a superoperator defined by
\begin{equation}
\mathcal{L}(t)\rho(t) = -\text{i}\left[H(t),\rho(t)\right]+\sum_k \mathcal{D}[L_k]\rho(t),
\end{equation} 
with the Dissipator defined as
\begin{equation}
\mathcal{D}[L]\rho(t) = \mathcal{J}[L]\rho(t) - \tfrac{1}{2}\{L^\dagger L,\rho(t)\}
\end{equation}
and the recycling (or emission) superoperator 
\begin{equation}
\mathcal{J}[L]\rho(t) = L \rho(t) L^\dagger.
\end{equation}
Finally, $L_k$ are the loss operators defined by the system operators $\sigma_k$ and their coupling rates to the reservoirs $\gamma_k$, i.e. $L_k=\sqrt{\gamma_k} \sigma_k$. Even though the $2X$ system physically emits into the same reservoirs for $\ket{2X}\rightarrow\ket{X}$ and $\ket{X}\rightarrow\ket{0}$, they are at such different frequencies ($E_b\gg \gamma,\gamma_X,\gamma_{2X}$) they can be considered to emit into separate Markovian reservoirs.

\begin{figure}[!t]
  \includegraphics[width=\columnwidth]{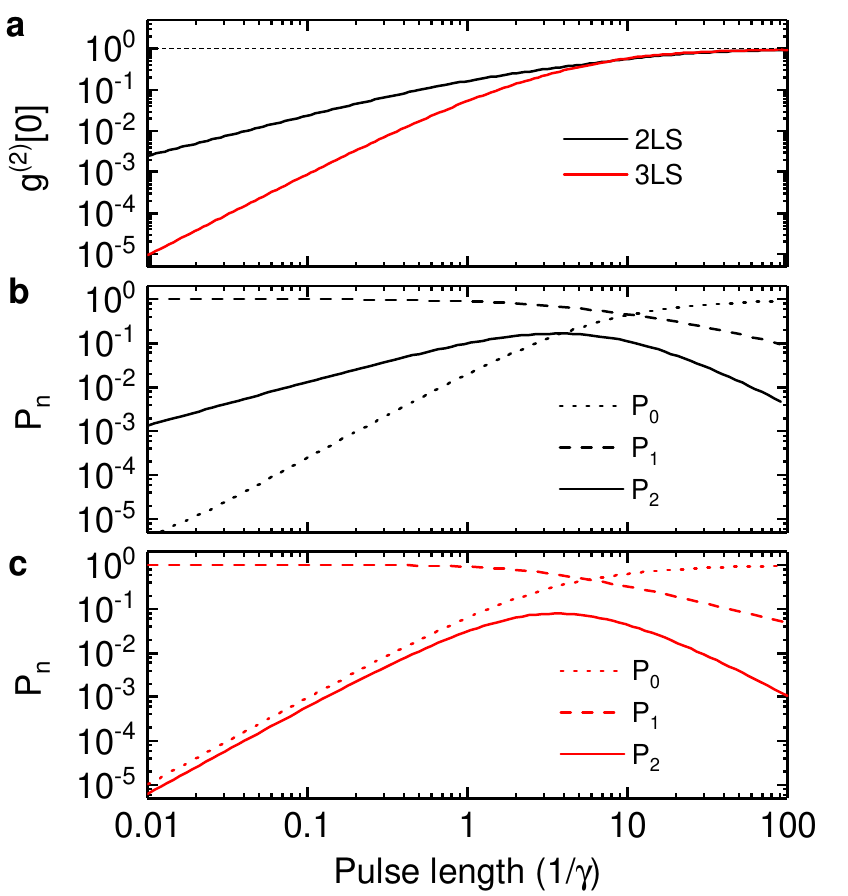}
  \caption{\textbf{Probabilities for different photodetection events.} (a) Simulated values of $g^{(2)}[0]$ as a function of the pulse length for a resonantly driven two-level system (black) and two-photon excitation of 2X (red). Dashed line represents Poissonian statistics of driving laser. (b-c) Simulated photocount distribution $P_n$, i.e. the probability for $n$ different detections to occur, either from the emission of (b) a 2LS or (c) a 3LS filtered on either the $2X$ or $X'$ transition frequency.
  }
  \label{figure:3}
\end{figure}

We can then calculate the pulse-wise second-order coherences $g^{(2)}_k[0]$ from the integrated versions of the correlators \cite{Fischer2016}
\begin{equation}
G^{(2)}_k(t_1,t_2)=\text{tr}\left[\mathcal{J}[L_k]\mathcal{V}(t_2,t_1)\mathcal{J}[L_k]\mathcal{V}(t_1,0)\rho(0)\right].
\end{equation}
These coherences were calculated for $A=\pi$ pulses (we took $\gamma=\gamma_{X}=\gamma_{2X}/2$), driving both the 2LS (Fig. \ref{figure:3}a---black) and 3LS (Fig. \ref{figure:3}a---red), and they very closely match the experimental results of Fig. \ref{figure:2}b. Small differences between experiment and theory result from experimental inaccuracies, such as the error in determining the pulse area of $\pi$, inaccuracies in the pulse shape, as well as drifts and fluctuations in power over the duration of the measurements \cite{Fischer2017-2}. We also note that we performed all quantum simulations with the Quantum Toolbox in Python (QuTiP) \cite{Johansson2014}.

The final goal is to compare these results to the theoretical photocount distribution $P_n$, from which the photodetectors sample. The ideal 2LS emits an entirely pure photonic state into the reservoir \cite{fischer2017scattering}, whereas the 3LS cascade is known to emit an entangled state between the reservoirs \cite{Fischer2016,Akopian2006}. From the perspective of a single reservoir, i.e. tracing over the other reservoir, this means the state could be highly mixed and hence our previous techniques would not apply easily \cite{Fischer2016, Fischer2017-2, fischer2017scattering}. Instead, we will use the Mandel photon counting formula, as connected to the system state by Carmichael \cite{carmichael2009open,wiseman2009quantum,gardiner2015quantum}. To do this, we first define a new superoperator
\begin{equation}
\mathcal{K}(t_1,t_0) = T_\leftarrow\text{exp}\left[\int_{t_0}^{t_1}\mathop{\text{d}t}\left(\mathcal{L}(t)-\mathcal{J}[L_k]\right)\right],
\end{equation}
which can be thought of as an unnormalized map that evolves the density matrix conditioned on no photon emissions into the $k$-th reservoir. Then, the total density matrix evolution can be unraveled with respect to $n$ emissions into the $k$-th reservoir as
\begin{eqnarray}
\rho(t) = \sum_{n=0}^\infty \int_0^{t}\mathop{\text{d}t_{n}}\int_0^{t_n}\mathop{\text{d}t_{n-1}}\cdots\int_0^{t_2}\mathop{\text{d}t_{1}} \mathcal{K}(t,t_n)\mathcal{J}[L_k]\times\nonumber\\
\mathcal{K}(t_n,t_{n-1})\mathcal{J}[L_k]\cdots\mathcal{K}(t_2,t_1)\mathcal{J}[L_k]\mathcal{K}(t_1,0)\rho(0).\nonumber\\
\end{eqnarray}
This summation is over different numbers of photon emissions into the $k$-th reservoir, such that the probability density for a sequence of $n$ emissions at times $t_1,t_2,\dots,t_n<t$ over the interval $[0,t]$ is given by
\begin{align}
p_n(t_1,t_2,\dots,t_n;[0,t])=&\nonumber\\
&\hspace{-15ex}\text{tr}\Big[\mathcal{K}(t,t_n)\mathcal{J}[L_k]\mathcal{K}(t_n,t_{n-1})\mathcal{J}[L_k]\cdots\nonumber\\
&\mathcal{K}(t_2,t_1)\mathcal{J}[L_k]\mathcal{K}(t_1,0)\rho(0)\Big].
\end{align} 
We drop the time label by taking the limit
\begin{eqnarray}
p_n(t_1,t_2,\dots,t_n)\equiv \lim_{t\rightarrow\infty}p_n(t_1,t_2,\dots,t_n;[0,t]),
\end{eqnarray}
which corresponds to the case where the system has entirely decayed after excitation by the laser pulse. In practice, we just integrate for a few spontaneous emission lifetimes after the pulse ends. Then, the photocount distribution into the $k$-th channel is given by
\begin{eqnarray}
P_n = \int_0^{\infty}\mathop{\text{d}t_{n}}\int_0^{t_n}\mathop{\text{d}t_{n-1}}\cdots\int_0^{t_2}\mathop{\text{d}t_{1}}p_n(t_1,t_2,\dots,t_n).\nonumber\\
\end{eqnarray}
Note there is only one possible way to count zero photons emitted:
\begin{equation}
P_0=\lim_{t\rightarrow\infty}\text{tr}\big[\mathcal{K}(t,0)\rho(0)\big].
\end{equation}

To our knowledge, this is the first use of such a model to extract photocount distributions for photon sources. For a single-photon source, driven with area $A=\pi$, only $P_n$ for $n<3$ are significant---we calculate these probabilities in Fig. \ref{figure:3} for (b) the 2LS and (c) the 3LS filtered on either transition frequency. For the single-photon sources, $g^{(2)}[0]\approx 2P_2/(P_1+2P_2)^2$, which scales linearly for the 2LS and quadratically for the 3LS with pulse length. Then, the error rate for the single-photon source is directly accessible as $P_2\approx g^{(2)}[0]/2$.

\section{Discussion}
As we have shown, the bi-excitonic and effective 3-level systems are superior single-photon sources to a two-level system. We briefly provide an approximate analysis which yields strong insight into the fundamental reason behind this behavior.

Previously, we derived an analytic estimate for the two-photon error rate $P_2$ and hence $g^{(2)}[0]$, for short pulses resonantly driving a 2LS \cite{Fischer2017-2}. Keeping only terms to first order in $\gamma T$, where $T$ is the pulse length, results for short pulses are $P_2 \approx \frac{\gamma T}{8}$ and $g^{(2)}[0] \approx \frac{\gamma T}{4}$. Here, we further derive an analytic estimate for the 3LS.

For the 3LS, the only way for two emissions to occur at the $2X$ transition frequency is through the cycle of transitions 
\begin{equation}
\ket{0}\rightarrow\ket{2X}\rightarrow\ket{X'}\rightarrow\ket{0}\rightarrow\ket{2X}.
\end{equation}
As in the two-level system, the first emission $\ket{2X}\rightarrow\ket{X'}$ contributes a factor of $\gamma_{2X}\sin(\frac{At}{2T})^2$ to the emission probability density, under the approximation of a square driving pulse. After the first emission, the system is almost exclusively in $\ket{X'}$ so $\ket{X'}\rightarrow\ket{0}$ contributes a factor of $\gamma_{X} \text{e}^{-\gamma_X t}$. Finally, $\ket{0}\rightarrow\ket{2X}$ again provides another $\gamma_{2X}\sin(\frac{At}{2T})^2$. Hence, the two-photon emission density into the first reservoir is approximately 
\begin{widetext}
\begin{subequations}
\begin{eqnarray}
p(t_1,t_2;t_1')&\approx&
\begin{cases}
\gamma_{2X}\sin(\frac{At_1}{2T})^2\gamma_X \text{e}^{-\gamma_X (t_1'-t_1)}\gamma_{2X}\sin(\frac{A(t_2-t_1')}{2T})^2& \text{ if } 0<t_1<t_1'<t_2<T\\
\gamma_{2X}\sin(\frac{At_1}{2T})^2\gamma_X \text{e}^{-\gamma_X (t_1'-t_1)}\gamma_{2X}\sin(\frac{A(T-t_1')}{2T})^2\text{e}^{-\gamma_{2X} (t_2-T)}& \text{ if } 0<t_1<t_1'<T<t_2 \\
0& \text{ otherwise }
\end{cases}\\
&\approx&
\begin{cases}
\gamma_{2X}\sin(\frac{At_1}{2T})^2\gamma_X \gamma_{2X}\sin(\frac{A(t_2-t_1')}{2T})^2& \text{ if } 0<t_1<t_1'<t_2<T\\
\gamma_{2X}\sin(\frac{At_1}{2T})^2\gamma_X \gamma_{2X}\sin(\frac{A(T-t_1')}{2T})^2\text{e}^{-\gamma_{2X} (t_2-T)}& \text{ if } 0<t_1<t_1'<T<t_2 \\
0& \text{ otherwise }
\end{cases}
\end{eqnarray}
\end{subequations}
\end{widetext}
where $t_1'$ is the time of emission of the first photon at the $X$ transition frequency. Hence, the two-photon error rate at the $2X$ frequency is 
\begin{equation}
P_2=\int_0^\infty\int_0^\infty\int_0^\infty \mathop{\text{d}t_1}\mathop{\text{d}t_1'}\mathop{\text{d}t_2} p(t_1,t_2;t_1').
\end{equation}
In the short pulse regime
\small
\begin{eqnarray}
P_2&\approx&\gamma_{2X}^2\gamma_{X}\int_0^T\int_{t_1}^T\int_{t_1'}^T \mathop{\text{d}t_1}\mathop{\text{d}t_1'}\mathop{\text{d}t_2}\sin(\frac{At_1}{2T})^2\sin(\frac{A(t_2-t_1')}{2T})^2\nonumber\\
&& + \gamma_{2X}\gamma_{X}\int_0^T\int_{t_1}^T \mathop{\text{d}t_1}\mathop{\text{d}t_1'}\sin(\frac{At_1}{2T})^2\sin(\frac{A(T-t_1')}{2T})^2\nonumber\\
&\approx&\mathcal{O}(T^4)+\mathcal{O}(T^2).
\end{eqnarray}
\normalsize
Since the leading order of the second integral is lowest, and for an $A=\pi$ pulse, the 3LS has
\begin{eqnarray}
P_2&\approx&\gamma_{2X}\gamma_{X}\int_0^T\int_{t_1}^T \mathop{\text{d}t_1}\mathop{\text{d}t_1'}\sin(\frac{At_1}{2T})^2\sin(\frac{A(T-t_1')}{2T})^2\nonumber\\
&\approx&\gamma_{2X}\gamma_{X} T^2 \frac{\pi^2-8}{8\pi^2}
\end{eqnarray}
or $g^{(2)}[0]\approx\gamma_{2X}\gamma_{X} T^2 (\pi^2-8)/4\pi^2$. This quadratic scaling matches exactly with the linear region of the theoretical results in Fig. 3 and the experimentally observed results in Fig. 2.

In summary,  we have quantified the re-excitation process in resonantly driven 2LS and two-photon excitation in 3LS. We have demonstrated that the 3LS, and hence the $2X$ system, dramatically suppresses re-excitation resulting in orders of magnitude better single-photon source operation. While our experiments are a proof-of-principle with a sample structure that does not provide a high collection efficiency, the technique is directly applicable to QDs embedded in nanophotonic structures where a very efficient detection of emitted photons is possible \cite{Gazzano2013, Claudon2010, Gschrey2015, Sapienza2015}. Beyond superior single-photon purity, the $2X$ scheme has the advantage over a resonantly driven 2LS that no cross-polarized suppression of the excitation laser is necessary because the driving laser and emission are far detuned such that the laser can easily be spectrally filtered out. Thus, it is easier to implement, as misalignment or optical imperfections do not pose an additional possibility for multi-photon errors. For example, very short laser pulses typically degrade cross-polarized suppression due to the increased spectral width of the laser, a wavelength sensitivity of the suppression and less efficient driving. Moreover, it does not reduce the source brightness which would be the case in a resonantly driven TLS where laser and signal are orthogonally polarized but need to couple to the same transition. To obtain high photon indistinguishability experimentally the emission has still to be filtered to one polarization, however Purcell enhancing just one polarization with a nanoresonator would transform the four-level system to an effective three-level system with a source brightness near unity. Therefore, we expect this scheme combined with appropriate nanoresonators be an excellent candidate for a single-photon source in future quantum information processors.

\section{Methods}
The sample investigated is grown by molecular beam epitaxy (MBE). It consists of a layer of InGaAs quantum dots with low areal density ($<1\mu m^{-2}$), embedded within the intrinsic region of a Schottky photodiode formed from an n-doped layer below the quantum dots and a semitransparent titanium gold front contact. The distance between the doped layer and the quantum dots is 35nm, which enables control over the charge status of the dot. A weak planar microcavity with an optical thickness of one wavelength is formed from a buried 18-pair GaAs/AlAs distributive Bragg reflector (DBR) and the semitransparent top contact, which enhances the in- and out-coupling of light.

All optical measurements were performed at 4.2K in a liquid helium dipstick set-up. For excitation and detection, a microscope objective with a numerical aperture of $NA=0.68$ was used. Cross-polarized measurements were performed using a polarizing beam splitter. To further enhance the extinction ratio, additional thin film linear polarizers were placed in the excitation/detection pathways and a single-mode fibre was used to spatially filter the detection signal. Furthermore, a quarter-wave plate was placed between the beamsplitter and the microscope objective to correct for birefringence of the optics and the sample itself.

For Figs. 1c and 1d, a  weak laser background (due to an imperfect suppression of the excitation laser) was subtracted. This linearly increasing background was directly measured through electrically tuning the quantum dot out of resonance, and typically amounted to less than $10\%$ of the signal by $5 \pi$ pulse area.

The 3ps to 80-ps-long excitation pulses were derived from a fs-pulsed titanium sapphire laser (Coherent Mira 900) through pulse shaping. For the 3ps to 25ps long pulses, a 4f pulse shaper with a focal length of 1m and an 1,800 lmm$^{-1}$ grating was used. For the 80ps long pulses a spectrometer-like filter with a focal length of 1m and an 1,800 lmm$^{-1}$ grating was used. Longer pulses were obtained through modulating a continuous wave laser. For the modulation, a fibre-coupled and EOM-controlled lithium niobate Mach-Zehnder (MZ) interferometer with a bandwidth of 10GHz (Photline NIR-MX-LN-10) was used. Such modulators allow control of the output intensity through a DC bias and a radiofrequency input. The radiofrequency pulses were generated by a 3.35GHz pulse-pattern generator (Agilent 81133A). To obtain a high extinction ratio, the temperature of the modulator was stabilized and precisely controlled (1mK) using a Peltier element, thermistor, and TEC controller. This enabled a static extinction ratio $>$45db.

Second-order autocorrelation measurements were performed using a Hanbury-Brown and Twiss (HBT) set-up consisting of one 50:50 beamsplitter and two single-photon avalanche diodes. The measured count rate for exciting $2X$ with a pulse of area $\pi$ was 9 kcps and the dark count rates of the detectors are $251\pm16$ cps and $95\pm10$ cps. The detected photons were correlated with a TimeHarp200 time-counting module. The time-bin width was 60 ps. The integration time for Fig. 2a was 11.05  hours and for Fig. 2b between 7.52 hours and 12.27 hours. In the pulse-wise form $g^{(2)}[0]=\frac{N_0}{N_1}$ where $N_0$ is the integrated area of the center peak and and $N_1$ is the average area of the side peaks. We used 16 side peaks for the averaging, which is the largest number that we can record with our correlation electronics and the used bin-width of 60ps. Note, that no long-term decay of the side peaks was observed, indicating the absence of any blinking and consistent with the fact that we use electronically stabilized devices. The error in $N_0$ and $N_1$ is given by $\sqrt{N_0}$ and $\sqrt{N_1}/4$, where the factor of 4 results from the fact that the integrated area of 16 peaks was used to calculate $N_1$. The error in $g^{(2)}[0]$ can then be calculated using quadratic propagation.

The measured correlations have a constant background that results from  dark counts of the detectors. To correct for the background, we first calculate the dark counts per time bin $n_{BG}$ by averaging a large number of bins between the peaks. The background corrected value of $g^{(2)}[0]$ is then given by $g^{(2)}_{\textrm{corr}}[0]=\frac{N_{0,\textrm{corr}}}{N_{1,\text{corr}}}$ where $N_{0,\textrm{corr}}=N_0-N_{\textrm{BG}}$ and $N_{1,\textrm{corr}}=N_1-N_{\textrm{BG}}$. Note, that the error still results from $\sqrt{N_0}$ and not $\sqrt{N_{0, \textrm{corr}}}$, highlighting the importance of detectors with low dark counts for the characterization single-photon sources with ultra-low multi photon error rates. 

\section{Data Availability}
The data that support the plots within this paper and other findings of this study are available from the corresponding authors upon reasonable request.

\section{Acknowledgements}
We gratefully acknowledge financial support from the DFG via the Nanosystems Initiative Munich, the BMBF via Q.Com (Project No. 16KIS0110), BaCaTeC, the International Graduate School of Science and Engineering (IGSSE) of TUM and the National Science Foundation (Division of Materials Research Grant No. 1503759). D.L. acknowledges support from the Fong Stanford Graduate Fellowship and the National Defense Science and Engineering Graduate Fellowship. J.W. acknowledges support from the PhD programme ExQM of the Elite Network of Bavaria. J.V. gratefully acknowledges support from the TUM Institute of Advanced Study. R.T. acknowledges support from the Kailath Stanford Graduate Fellowship. K.M acknowledges support from the Bavarian Academy of Sciences and Humanities.

\section{Competing Interests}
The authors declare no competing financial and non-financial interests.

\section{Author Contributions}
L.H., S.A., J.W. and K.M. performed the experiments. K.A.F., D.L., S.S. and R.T.  performed the theoretical work and modelling. J.V. and J.J.F. provided expertise. K.A.F. and K.M. conceived the idea. All authors participated in the discussion and understanding of the results. L.H. and K.A.F. contributed equally.

\bibliography{BibliographyX}

\providecommand{\noopsort}[1]{}\providecommand{\singleletter}[1]{#1}%
\begin{thebibliography}{37}%
\makeatletter
\providecommand \@ifxundefined [1]{%
 \@ifx{#1\undefined}
}%
\providecommand \@ifnum [1]{%
 \ifnum #1\expandafter \@firstoftwo
 \else \expandafter \@secondoftwo
 \fi
}%
\providecommand \@ifx [1]{%
 \ifx #1\expandafter \@firstoftwo
 \else \expandafter \@secondoftwo
 \fi
}%
\providecommand \natexlab [1]{#1}%
\providecommand \enquote  [1]{``#1''}%
\providecommand \bibnamefont  [1]{#1}%
\providecommand \bibfnamefont [1]{#1}%
\providecommand \citenamefont [1]{#1}%
\providecommand \href@noop [0]{\@secondoftwo}%
\providecommand \href [0]{\begingroup \@sanitize@url \@href}%
\providecommand \@href[1]{\@@startlink{#1}\@@href}%
\providecommand \@@href[1]{\endgroup#1\@@endlink}%
\providecommand \@sanitize@url [0]{\catcode `\\12\catcode `\$12\catcode
  `\&12\catcode `\#12\catcode `\^12\catcode `\_12\catcode `\%12\relax}%
\providecommand \@@startlink[1]{}%
\providecommand \@@endlink[0]{}%
\providecommand \url  [0]{\begingroup\@sanitize@url \@url }%
\providecommand \@url [1]{\endgroup\@href {#1}{\urlprefix }}%
\providecommand \urlprefix  [0]{URL }%
\providecommand \Eprint [0]{\href }%
\providecommand \doibase [0]{http://dx.doi.org/}%
\providecommand \selectlanguage [0]{\@gobble}%
\providecommand \bibinfo  [0]{\@secondoftwo}%
\providecommand \bibfield  [0]{\@secondoftwo}%
\providecommand \translation [1]{[#1]}%
\providecommand \BibitemOpen [0]{}%
\providecommand \bibitemStop [0]{}%
\providecommand \bibitemNoStop [0]{.\EOS\space}%
\providecommand \EOS [0]{\spacefactor3000\relax}%
\providecommand \BibitemShut  [1]{\csname bibitem#1\endcsname}%
\let\auto@bib@innerbib\@empty
\bibitem [{\citenamefont {Michler}(2017)}]{michler}%
  \BibitemOpen
  \bibinfo {editor} {\bibfnamefont {Peter}\ \bibnamefont {Michler}},\ ed.,\
  \href@noop {} {\emph {\bibinfo {title} {Quantum Dots for Quantum Information
  Technologies}}}\ (\bibinfo  {publisher} {Springer},\ \bibinfo {year}
  {2017})\BibitemShut {NoStop}%
\bibitem [{\citenamefont {Senellart}\ \emph {et~al.}(2017)\citenamefont
  {Senellart}, \citenamefont {Solomon},\ and\ \citenamefont
  {White}}]{Senellart2017}%
  \BibitemOpen
  \bibfield  {author} {\bibinfo {author} {\bibfnamefont {Pascale}\ \bibnamefont
  {Senellart}}, \bibinfo {author} {\bibfnamefont {Glenn}\ \bibnamefont
  {Solomon}}, \ and\ \bibinfo {author} {\bibfnamefont {Andrew}\ \bibnamefont
  {White}},\ }\bibfield  {title} {\enquote {\bibinfo {title} {{High-performance
  semiconductor quantum-dot single-photon sources}},}\ }\href {\doibase
  10.1038/nnano.2017.218} {\bibfield  {journal} {\bibinfo  {journal} {Nature
  Nanotechnology}\ }\textbf {\bibinfo {volume} {12}},\ \bibinfo {pages}
  {1026--1039} (\bibinfo {year} {2017})}\BibitemShut {NoStop}%
\bibitem [{\citenamefont {Prechtel}\ \emph {et~al.}(2013)\citenamefont
  {Prechtel}, \citenamefont {Kuhlmann}, \citenamefont {Houel}, \citenamefont
  {Greuter}, \citenamefont {Ludwig}, \citenamefont {Reuter}, \citenamefont
  {Wieck},\ and\ \citenamefont {Warburton}}]{Prechtel2013}%
  \BibitemOpen
  \bibfield  {author} {\bibinfo {author} {\bibfnamefont {Jonathan~H}\
  \bibnamefont {Prechtel}}, \bibinfo {author} {\bibfnamefont {Andreas~V}\
  \bibnamefont {Kuhlmann}}, \bibinfo {author} {\bibfnamefont {Julien}\
  \bibnamefont {Houel}}, \bibinfo {author} {\bibfnamefont {Lukas}\ \bibnamefont
  {Greuter}}, \bibinfo {author} {\bibfnamefont {Arne}\ \bibnamefont {Ludwig}},
  \bibinfo {author} {\bibfnamefont {Dirk}\ \bibnamefont {Reuter}}, \bibinfo
  {author} {\bibfnamefont {Andreas~D}\ \bibnamefont {Wieck}}, \ and\ \bibinfo
  {author} {\bibfnamefont {Richard~J}\ \bibnamefont {Warburton}},\ }\bibfield
  {title} {\enquote {\bibinfo {title} {{Frequency-stabilized source of single
  photons from a solid-state qubit}},}\ }\href {\doibase
  10.1103/PhysRevX.3.04106} {\bibfield  {journal} {\bibinfo  {journal}
  {Physical Review X}\ }\textbf {\bibinfo {volume} {3}},\ \bibinfo {pages}
  {041006} (\bibinfo {year} {2013})}\BibitemShut {NoStop}%
\bibitem [{\citenamefont {Hansom}\ \emph {et~al.}(2014)\citenamefont {Hansom},
  \citenamefont {Schulte}, \citenamefont {Matthiesen}, \citenamefont
  {Stanley},\ and\ \citenamefont {Atature}}]{Hansom2014}%
  \BibitemOpen
  \bibfield  {author} {\bibinfo {author} {\bibfnamefont {Jack}\ \bibnamefont
  {Hansom}}, \bibinfo {author} {\bibfnamefont {Carsten~H.H.}\ \bibnamefont
  {Schulte}}, \bibinfo {author} {\bibfnamefont {Clemens}\ \bibnamefont
  {Matthiesen}}, \bibinfo {author} {\bibfnamefont {Megan~J}\ \bibnamefont
  {Stanley}}, \ and\ \bibinfo {author} {\bibfnamefont {Mete}\ \bibnamefont
  {Atature}},\ }\bibfield  {title} {\enquote {\bibinfo {title} {{Frequency
  stabilization of the zero-phonon line of a quantum dot via phonon-assisted
  active feedback}},}\ }\href {\doibase 10.1063/1.4901045} {\bibfield
  {journal} {\bibinfo  {journal} {Applied Physics Letters}\ }\textbf {\bibinfo
  {volume} {105}},\ \bibinfo {pages} {172107} (\bibinfo {year}
  {2014})}\BibitemShut {NoStop}%
\bibitem [{\citenamefont {Kuhlmann}\ \emph {et~al.}(2015)\citenamefont
  {Kuhlmann}, \citenamefont {Prechtel}, \citenamefont {Houel}, \citenamefont
  {Ludwig}, \citenamefont {Reuter}, \citenamefont {Wieck},\ and\ \citenamefont
  {Warburton}}]{Kuhlmann2015}%
  \BibitemOpen
  \bibfield  {author} {\bibinfo {author} {\bibfnamefont {Andreas~V.}\
  \bibnamefont {Kuhlmann}}, \bibinfo {author} {\bibfnamefont {Jonathan~H.}\
  \bibnamefont {Prechtel}}, \bibinfo {author} {\bibfnamefont {Julien}\
  \bibnamefont {Houel}}, \bibinfo {author} {\bibfnamefont {Arne}\ \bibnamefont
  {Ludwig}}, \bibinfo {author} {\bibfnamefont {Dirk}\ \bibnamefont {Reuter}},
  \bibinfo {author} {\bibfnamefont {Andreas~D.}\ \bibnamefont {Wieck}}, \ and\
  \bibinfo {author} {\bibfnamefont {Richard~J.}\ \bibnamefont {Warburton}},\
  }\bibfield  {title} {\enquote {\bibinfo {title} {{Transform-limited single
  photons from a single quantum dot}},}\ }\href {\doibase 10.1038/ncomms9204}
  {\bibfield  {journal} {\bibinfo  {journal} {Nature Communications}\ }\textbf
  {\bibinfo {volume} {6}},\ \bibinfo {pages} {8204} (\bibinfo {year}
  {2015})}\BibitemShut {NoStop}%
\bibitem [{\citenamefont {He}\ \emph {et~al.}(2013)\citenamefont {He},
  \citenamefont {He}, \citenamefont {Wei}, \citenamefont {Wu}, \citenamefont
  {Atature}, \citenamefont {Schneider}, \citenamefont {Hofling}, \citenamefont
  {Kamp}, \citenamefont {Lu},\ and\ \citenamefont {Pan}}]{He2013}%
  \BibitemOpen
  \bibfield  {author} {\bibinfo {author} {\bibfnamefont {Yu-Ming}\ \bibnamefont
  {He}}, \bibinfo {author} {\bibfnamefont {Yu}~\bibnamefont {He}}, \bibinfo
  {author} {\bibfnamefont {Yu-Jia}\ \bibnamefont {Wei}}, \bibinfo {author}
  {\bibfnamefont {Dian}\ \bibnamefont {Wu}}, \bibinfo {author} {\bibfnamefont
  {Mete}\ \bibnamefont {Atature}}, \bibinfo {author} {\bibfnamefont
  {Christian}\ \bibnamefont {Schneider}}, \bibinfo {author} {\bibfnamefont
  {Sven}\ \bibnamefont {Hofling}}, \bibinfo {author} {\bibfnamefont {Martin}\
  \bibnamefont {Kamp}}, \bibinfo {author} {\bibfnamefont {Chao-Yang}\
  \bibnamefont {Lu}}, \ and\ \bibinfo {author} {\bibfnamefont {Jian-Wei}\
  \bibnamefont {Pan}},\ }\bibfield  {title} {\enquote {\bibinfo {title}
  {{On-demand semiconductor single-photon source with near-unity
  indistinguishability}},}\ }\href {\doibase 10.1038/nnano.2012.262} {\bibfield
   {journal} {\bibinfo  {journal} {Nature Nanotechnology}\ }\textbf {\bibinfo
  {volume} {8}},\ \bibinfo {pages} {213--217} (\bibinfo {year}
  {2013})}\BibitemShut {NoStop}%
\bibitem [{\citenamefont {Somaschi}\ \emph {et~al.}(2016)\citenamefont
  {Somaschi}, \citenamefont {Giesz}, \citenamefont {De~Santis}, \citenamefont
  {Loredo}, \citenamefont {Almeida}, \citenamefont {Hornecker}, \citenamefont
  {Portalupi}, \citenamefont {Grange}, \citenamefont {Ant{\'o}n}, \citenamefont
  {Demory}, \citenamefont {G{\'o}mez}, \citenamefont {Sagnes}, \citenamefont
  {Lanzillotti-Kimura}, \citenamefont {Lema{\'\i}tre}, \citenamefont
  {Auffeves}, \citenamefont {White}, \citenamefont {Lanco},\ and\ \citenamefont
  {Senellart}}]{Somaschi2016}%
  \BibitemOpen
  \bibfield  {author} {\bibinfo {author} {\bibfnamefont {N.}~\bibnamefont
  {Somaschi}}, \bibinfo {author} {\bibfnamefont {V.}~\bibnamefont {Giesz}},
  \bibinfo {author} {\bibfnamefont {L.}~\bibnamefont {De~Santis}}, \bibinfo
  {author} {\bibfnamefont {J.~C.}\ \bibnamefont {Loredo}}, \bibinfo {author}
  {\bibfnamefont {M.~P.}\ \bibnamefont {Almeida}}, \bibinfo {author}
  {\bibfnamefont {G.}~\bibnamefont {Hornecker}}, \bibinfo {author}
  {\bibfnamefont {S.~L.}\ \bibnamefont {Portalupi}}, \bibinfo {author}
  {\bibfnamefont {T.}~\bibnamefont {Grange}}, \bibinfo {author} {\bibfnamefont
  {C.}~\bibnamefont {Ant{\'o}n}}, \bibinfo {author} {\bibfnamefont
  {J.}~\bibnamefont {Demory}}, \bibinfo {author} {\bibfnamefont
  {C.}~\bibnamefont {G{\'o}mez}}, \bibinfo {author} {\bibfnamefont
  {I.}~\bibnamefont {Sagnes}}, \bibinfo {author} {\bibfnamefont {N.~D.}\
  \bibnamefont {Lanzillotti-Kimura}}, \bibinfo {author} {\bibfnamefont
  {A.}~\bibnamefont {Lema{\'\i}tre}}, \bibinfo {author} {\bibfnamefont
  {A.}~\bibnamefont {Auffeves}}, \bibinfo {author} {\bibfnamefont {A.~G.}\
  \bibnamefont {White}}, \bibinfo {author} {\bibfnamefont {L.}~\bibnamefont
  {Lanco}}, \ and\ \bibinfo {author} {\bibfnamefont {P.}~\bibnamefont
  {Senellart}},\ }\bibfield  {title} {\enquote {\bibinfo {title} {Near-optimal
  single-photon sources in the solid state},}\ }\href
  {http://dx.doi.org/10.1038/nphoton.2016.23} {\bibfield  {journal} {\bibinfo
  {journal} {Nature Photonics}\ }\textbf {\bibinfo {volume} {10}},\ \bibinfo
  {pages} {340--345} (\bibinfo {year} {2016})}\BibitemShut {NoStop}%
\bibitem [{\citenamefont {Unsleber}\ \emph {et~al.}(2016)\citenamefont
  {Unsleber}, \citenamefont {He}, \citenamefont {Gerhardt}, \citenamefont
  {Maier}, \citenamefont {Lu}, \citenamefont {Pan}, \citenamefont {Gregersen},
  \citenamefont {Kamp}, \citenamefont {Schneider},\ and\ \citenamefont
  {H\"{o}fling}}]{Unsleber2016}%
  \BibitemOpen
  \bibfield  {author} {\bibinfo {author} {\bibfnamefont {Sebastian}\
  \bibnamefont {Unsleber}}, \bibinfo {author} {\bibfnamefont {Yu-Ming}\
  \bibnamefont {He}}, \bibinfo {author} {\bibfnamefont {Stefan}\ \bibnamefont
  {Gerhardt}}, \bibinfo {author} {\bibfnamefont {Sebastian}\ \bibnamefont
  {Maier}}, \bibinfo {author} {\bibfnamefont {Chao-Yang}\ \bibnamefont {Lu}},
  \bibinfo {author} {\bibfnamefont {Jian-Wei}\ \bibnamefont {Pan}}, \bibinfo
  {author} {\bibfnamefont {Niels}\ \bibnamefont {Gregersen}}, \bibinfo {author}
  {\bibfnamefont {Martin}\ \bibnamefont {Kamp}}, \bibinfo {author}
  {\bibfnamefont {Christian}\ \bibnamefont {Schneider}}, \ and\ \bibinfo
  {author} {\bibfnamefont {Sven}\ \bibnamefont {H\"{o}fling}},\ }\bibfield
  {title} {\enquote {\bibinfo {title} {Highly indistinguishable on-demand
  resonance fluorescence photons from a deterministic quantum dot micropillar
  device with 74 \% extraction efficiency},}\ }\href {\doibase
  10.1364/OE.24.008539} {\bibfield  {journal} {\bibinfo  {journal} {Optics
  Express}\ }\textbf {\bibinfo {volume} {24}},\ \bibinfo {pages} {8539--8546}
  (\bibinfo {year} {2016})}\BibitemShut {NoStop}%
\bibitem [{\citenamefont {Ding}\ \emph {et~al.}(2016)\citenamefont {Ding},
  \citenamefont {He}, \citenamefont {Duan}, \citenamefont {Gregersen},
  \citenamefont {Chen}, \citenamefont {Unsleber}, \citenamefont {Maier},
  \citenamefont {Schneider}, \citenamefont {Kamp}, \citenamefont {H\"ofling},
  \citenamefont {Lu},\ and\ \citenamefont {Pan}}]{Ding2016}%
  \BibitemOpen
  \bibfield  {author} {\bibinfo {author} {\bibfnamefont {Xing}\ \bibnamefont
  {Ding}}, \bibinfo {author} {\bibfnamefont {Yu}~\bibnamefont {He}}, \bibinfo
  {author} {\bibfnamefont {Z.-C.}\ \bibnamefont {Duan}}, \bibinfo {author}
  {\bibfnamefont {Niels}\ \bibnamefont {Gregersen}}, \bibinfo {author}
  {\bibfnamefont {M.-C.}\ \bibnamefont {Chen}}, \bibinfo {author}
  {\bibfnamefont {S.}~\bibnamefont {Unsleber}}, \bibinfo {author}
  {\bibfnamefont {S.}~\bibnamefont {Maier}}, \bibinfo {author} {\bibfnamefont
  {Christian}\ \bibnamefont {Schneider}}, \bibinfo {author} {\bibfnamefont
  {Martin}\ \bibnamefont {Kamp}}, \bibinfo {author} {\bibfnamefont {Sven}\
  \bibnamefont {H\"ofling}}, \bibinfo {author} {\bibfnamefont {Chao-Yang}\
  \bibnamefont {Lu}}, \ and\ \bibinfo {author} {\bibfnamefont {Jian-Wei}\
  \bibnamefont {Pan}},\ }\bibfield  {title} {\enquote {\bibinfo {title}
  {On-demand single photons with high extraction efficiency and near-unity
  indistinguishability from a resonantly driven quantum dot in a
  micropillar},}\ }\href {\doibase 10.1103/PhysRevLett.116.020401} {\bibfield
  {journal} {\bibinfo  {journal} {Physical Review Letters}\ }\textbf {\bibinfo
  {volume} {116}},\ \bibinfo {pages} {020401} (\bibinfo {year}
  {2016})}\BibitemShut {NoStop}%
\bibitem [{\citenamefont {He}\ \emph {et~al.}(2016)\citenamefont {He},
  \citenamefont {Liu}, \citenamefont {Maier}, \citenamefont {Emmerling},
  \citenamefont {Gerhardt}, \citenamefont {Davanco}, \citenamefont
  {Srinivasan}, \citenamefont {Schneider},\ and\ \citenamefont
  {H{\"{o}}fling}}]{He2016}%
  \BibitemOpen
  \bibfield  {author} {\bibinfo {author} {\bibfnamefont {Yu-Ming}\ \bibnamefont
  {He}}, \bibinfo {author} {\bibfnamefont {Jin}\ \bibnamefont {Liu}}, \bibinfo
  {author} {\bibfnamefont {Sebastian}\ \bibnamefont {Maier}}, \bibinfo {author}
  {\bibfnamefont {Monika}\ \bibnamefont {Emmerling}}, \bibinfo {author}
  {\bibfnamefont {Stefan}\ \bibnamefont {Gerhardt}}, \bibinfo {author}
  {\bibfnamefont {Marcelo}\ \bibnamefont {Davanco}}, \bibinfo {author}
  {\bibfnamefont {Kartik}\ \bibnamefont {Srinivasan}}, \bibinfo {author}
  {\bibfnamefont {Christian}\ \bibnamefont {Schneider}}, \ and\ \bibinfo
  {author} {\bibfnamefont {Sven}\ \bibnamefont {H{\"{o}}fling}},\ }\bibfield
  {title} {\enquote {\bibinfo {title} {{Deterministic implementation of a
  bright, on-demand single photon source with near-unity indistinguishability
  via quantum dot imaging}},}\ }\href {\doibase 10.1364/OPTICA.4.000802}
  {\bibfield  {journal} {\bibinfo  {journal} {Optica}\ }\textbf {\bibinfo
  {volume} {4}},\ \bibinfo {pages} {802} (\bibinfo {year} {2016})}\BibitemShut
  {NoStop}%
\bibitem [{\citenamefont {M{\"{u}}ller}\ \emph {et~al.}(2016)\citenamefont
  {M{\"{u}}ller}, \citenamefont {Fischer}, \citenamefont {Dory}, \citenamefont
  {Sarmiento}, \citenamefont {Lagoudakis}, \citenamefont {Rundquist},
  \citenamefont {Kelaita},\ and\ \citenamefont
  {Vu{\v{c}}kovi{\'{c}}}}]{Mueller2016}%
  \BibitemOpen
  \bibfield  {author} {\bibinfo {author} {\bibfnamefont {Kai}\ \bibnamefont
  {M{\"{u}}ller}}, \bibinfo {author} {\bibfnamefont {Kevin~A.}\ \bibnamefont
  {Fischer}}, \bibinfo {author} {\bibfnamefont {Constantin}\ \bibnamefont
  {Dory}}, \bibinfo {author} {\bibfnamefont {Tomas}\ \bibnamefont {Sarmiento}},
  \bibinfo {author} {\bibfnamefont {Konstantinos~G.}\ \bibnamefont
  {Lagoudakis}}, \bibinfo {author} {\bibfnamefont {Armand}\ \bibnamefont
  {Rundquist}}, \bibinfo {author} {\bibfnamefont {Yousif~A.}\ \bibnamefont
  {Kelaita}}, \ and\ \bibinfo {author} {\bibfnamefont {Jelena}\ \bibnamefont
  {Vu{\v{c}}kovi{\'{c}}}},\ }\bibfield  {title} {\enquote {\bibinfo {title}
  {{Self-homodyne-enabled generation of indistinguishable photons}},}\ }\href
  {\doibase 10.1364/OPTICA.3.000931} {\bibfield  {journal} {\bibinfo  {journal}
  {Optica}\ }\textbf {\bibinfo {volume} {3}},\ \bibinfo {pages} {931} (\bibinfo
  {year} {2016})}\BibitemShut {NoStop}%
\bibitem [{\citenamefont {Wang}\ \emph {et~al.}(2016)\citenamefont {Wang},
  \citenamefont {Duan}, \citenamefont {Li}, \citenamefont {Chen}, \citenamefont
  {Li}, \citenamefont {He}, \citenamefont {Chen}, \citenamefont {He},
  \citenamefont {Ding}, \citenamefont {Peng}, \citenamefont {Schneider},
  \citenamefont {Kamp}, \citenamefont {H{\"{o}}fling}, \citenamefont {Lu},\
  and\ \citenamefont {Pan}}]{Wang2016}%
  \BibitemOpen
  \bibfield  {author} {\bibinfo {author} {\bibfnamefont {Hui}\ \bibnamefont
  {Wang}}, \bibinfo {author} {\bibfnamefont {Z.~C.}\ \bibnamefont {Duan}},
  \bibinfo {author} {\bibfnamefont {Y.~H.}\ \bibnamefont {Li}}, \bibinfo
  {author} {\bibfnamefont {Si}~\bibnamefont {Chen}}, \bibinfo {author}
  {\bibfnamefont {J.~P.}\ \bibnamefont {Li}}, \bibinfo {author} {\bibfnamefont
  {Y.~M.}\ \bibnamefont {He}}, \bibinfo {author} {\bibfnamefont {M.~C.}\
  \bibnamefont {Chen}}, \bibinfo {author} {\bibfnamefont {Yu}~\bibnamefont
  {He}}, \bibinfo {author} {\bibfnamefont {X.}~\bibnamefont {Ding}}, \bibinfo
  {author} {\bibfnamefont {Cheng~Zhi}\ \bibnamefont {Peng}}, \bibinfo {author}
  {\bibfnamefont {Christian}\ \bibnamefont {Schneider}}, \bibinfo {author}
  {\bibfnamefont {Martin}\ \bibnamefont {Kamp}}, \bibinfo {author}
  {\bibfnamefont {Sven}\ \bibnamefont {H{\"{o}}fling}}, \bibinfo {author}
  {\bibfnamefont {Chao~Yang}\ \bibnamefont {Lu}}, \ and\ \bibinfo {author}
  {\bibfnamefont {Jian~Wei}\ \bibnamefont {Pan}},\ }\bibfield  {title}
  {\enquote {\bibinfo {title} {{Near-Transform-Limited Single Photons from an
  Efficient Solid-State Quantum Emitter}},}\ }\href {\doibase
  10.1103/PhysRevLett.116.213601} {\bibfield  {journal} {\bibinfo  {journal}
  {Physical Review Letters}\ }\textbf {\bibinfo {volume} {116}},\ \bibinfo
  {pages} {213601} (\bibinfo {year} {2016})}\BibitemShut {NoStop}%
\bibitem [{\citenamefont {Loredo}\ \emph {et~al.}(2016)\citenamefont {Loredo},
  \citenamefont {Zakaria}, \citenamefont {Somaschi}, \citenamefont {Anton},
  \citenamefont {de~Santis}, \citenamefont {Giesz}, \citenamefont {Grange},
  \citenamefont {Broome}, \citenamefont {Gazzano}, \citenamefont {Coppola},
  \citenamefont {Sagnes}, \citenamefont {Lemaitre}, \citenamefont {Auffeves},
  \citenamefont {Senellart}, \citenamefont {Almeida},\ and\ \citenamefont
  {White}}]{Loredo2016}%
  \BibitemOpen
  \bibfield  {author} {\bibinfo {author} {\bibfnamefont {Juan~C.}\ \bibnamefont
  {Loredo}}, \bibinfo {author} {\bibfnamefont {Nor~A.}\ \bibnamefont
  {Zakaria}}, \bibinfo {author} {\bibfnamefont {Niccolo}\ \bibnamefont
  {Somaschi}}, \bibinfo {author} {\bibfnamefont {Carlos}\ \bibnamefont
  {Anton}}, \bibinfo {author} {\bibfnamefont {Lorenzo}\ \bibnamefont
  {de~Santis}}, \bibinfo {author} {\bibfnamefont {Valerian}\ \bibnamefont
  {Giesz}}, \bibinfo {author} {\bibfnamefont {Thomas}\ \bibnamefont {Grange}},
  \bibinfo {author} {\bibfnamefont {Matthew~A.}\ \bibnamefont {Broome}},
  \bibinfo {author} {\bibfnamefont {Olivier}\ \bibnamefont {Gazzano}}, \bibinfo
  {author} {\bibfnamefont {Guillaume}\ \bibnamefont {Coppola}}, \bibinfo
  {author} {\bibfnamefont {Isabelle}\ \bibnamefont {Sagnes}}, \bibinfo {author}
  {\bibfnamefont {Aristide}\ \bibnamefont {Lemaitre}}, \bibinfo {author}
  {\bibfnamefont {Alexia}\ \bibnamefont {Auffeves}}, \bibinfo {author}
  {\bibfnamefont {Pascale}\ \bibnamefont {Senellart}}, \bibinfo {author}
  {\bibfnamefont {Marcelo~P.}\ \bibnamefont {Almeida}}, \ and\ \bibinfo
  {author} {\bibfnamefont {Andrew~G.}\ \bibnamefont {White}},\ }\bibfield
  {title} {\enquote {\bibinfo {title} {{Scalable performance in solid-state
  single-photon sources}},}\ }\href {\doibase 10.1364/OPTICA.3.000433}
  {\bibfield  {journal} {\bibinfo  {journal} {Optica}\ }\textbf {\bibinfo
  {volume} {3}},\ \bibinfo {pages} {433} (\bibinfo {year} {2016})}\BibitemShut
  {NoStop}%
\bibitem [{\citenamefont {Iles-Smith}\ \emph {et~al.}(2017)\citenamefont
  {Iles-Smith}, \citenamefont {McCutcheon}, \citenamefont {Nazir},\ and\
  \citenamefont {M{\o}rk}}]{Iles-Smith2017}%
  \BibitemOpen
  \bibfield  {author} {\bibinfo {author} {\bibfnamefont {Jake}\ \bibnamefont
  {Iles-Smith}}, \bibinfo {author} {\bibfnamefont {Dara P.~S.}\ \bibnamefont
  {McCutcheon}}, \bibinfo {author} {\bibfnamefont {Ahsan}\ \bibnamefont
  {Nazir}}, \ and\ \bibinfo {author} {\bibfnamefont {Jesper}\ \bibnamefont
  {M{\o}rk}},\ }\bibfield  {title} {\enquote {\bibinfo {title} {{Phonon
  scattering inhibits simultaneous near-unity efficiency and
  indistinguishability in semiconductor single-photon sources}},}\ }\href
  {\doibase 10.1038/nphoton.2017.101} {\bibfield  {journal} {\bibinfo
  {journal} {Nature Photonics}\ }\textbf {\bibinfo {volume} {11}},\ \bibinfo
  {pages} {521--526} (\bibinfo {year} {2017})}\BibitemShut {NoStop}%
\bibitem [{\citenamefont {Loredo}\ \emph {et~al.}(2017)\citenamefont {Loredo},
  \citenamefont {Broome}, \citenamefont {Hilaire}, \citenamefont {Gazzano},
  \citenamefont {Sagnes}, \citenamefont {Lemaitre}, \citenamefont {Almeida},
  \citenamefont {Senellart},\ and\ \citenamefont {White}}]{Loredo2017}%
  \BibitemOpen
  \bibfield  {author} {\bibinfo {author} {\bibfnamefont {J.~C.}\ \bibnamefont
  {Loredo}}, \bibinfo {author} {\bibfnamefont {M.~A.}\ \bibnamefont {Broome}},
  \bibinfo {author} {\bibfnamefont {P.}~\bibnamefont {Hilaire}}, \bibinfo
  {author} {\bibfnamefont {O.}~\bibnamefont {Gazzano}}, \bibinfo {author}
  {\bibfnamefont {I.}~\bibnamefont {Sagnes}}, \bibinfo {author} {\bibfnamefont
  {A.}~\bibnamefont {Lemaitre}}, \bibinfo {author} {\bibfnamefont {M.~P.}\
  \bibnamefont {Almeida}}, \bibinfo {author} {\bibfnamefont {P.}~\bibnamefont
  {Senellart}}, \ and\ \bibinfo {author} {\bibfnamefont {A.~G.}\ \bibnamefont
  {White}},\ }\bibfield  {title} {\enquote {\bibinfo {title} {{Boson Sampling
  with Single-Photon Fock States from a Bright Solid-State Source}},}\ }\href
  {\doibase 10.1103/PhysRevLett.118.130503} {\bibfield  {journal} {\bibinfo
  {journal} {Physical Review Letters}\ }\textbf {\bibinfo {volume} {118}},\
  \bibinfo {pages} {130503} (\bibinfo {year} {2017})}\BibitemShut {NoStop}%
\bibitem [{\citenamefont {Wang}\ \emph {et~al.}(2017)\citenamefont {Wang},
  \citenamefont {He}, \citenamefont {Li}, \citenamefont {Su}, \citenamefont
  {Li}, \citenamefont {Huang}, \citenamefont {Ding}, \citenamefont {Chen},
  \citenamefont {Liu}, \citenamefont {Qin}, \citenamefont {Li}, \citenamefont
  {He}, \citenamefont {Schneider}, \citenamefont {Kamp}, \citenamefont {Peng},
  \citenamefont {H{\"{o}}fling}, \citenamefont {Lu},\ and\ \citenamefont
  {Pan}}]{Wang2017}%
  \BibitemOpen
  \bibfield  {author} {\bibinfo {author} {\bibfnamefont {Hui}\ \bibnamefont
  {Wang}}, \bibinfo {author} {\bibfnamefont {Yu}~\bibnamefont {He}}, \bibinfo
  {author} {\bibfnamefont {Yu-Huai}\ \bibnamefont {Li}}, \bibinfo {author}
  {\bibfnamefont {Zu-En}\ \bibnamefont {Su}}, \bibinfo {author} {\bibfnamefont
  {Bo}~\bibnamefont {Li}}, \bibinfo {author} {\bibfnamefont {He-Liang}\
  \bibnamefont {Huang}}, \bibinfo {author} {\bibfnamefont {Xing}\ \bibnamefont
  {Ding}}, \bibinfo {author} {\bibfnamefont {Ming-Cheng}\ \bibnamefont {Chen}},
  \bibinfo {author} {\bibfnamefont {Chang}\ \bibnamefont {Liu}}, \bibinfo
  {author} {\bibfnamefont {Jian}\ \bibnamefont {Qin}}, \bibinfo {author}
  {\bibfnamefont {Jin-Peng}\ \bibnamefont {Li}}, \bibinfo {author}
  {\bibfnamefont {Yu-Ming}\ \bibnamefont {He}}, \bibinfo {author}
  {\bibfnamefont {Christian}\ \bibnamefont {Schneider}}, \bibinfo {author}
  {\bibfnamefont {Martin}\ \bibnamefont {Kamp}}, \bibinfo {author}
  {\bibfnamefont {Cheng-Zhi}\ \bibnamefont {Peng}}, \bibinfo {author}
  {\bibfnamefont {Sven}\ \bibnamefont {H{\"{o}}fling}}, \bibinfo {author}
  {\bibfnamefont {Chao-Yang}\ \bibnamefont {Lu}}, \ and\ \bibinfo {author}
  {\bibfnamefont {Jian-Wei}\ \bibnamefont {Pan}},\ }\bibfield  {title}
  {\enquote {\bibinfo {title} {{High-efficiency multiphoton boson sampling}},}\
  }\href {\doibase 10.1038/nphoton.2017.63} {\bibfield  {journal} {\bibinfo
  {journal} {Nature Photonics}\ }\textbf {\bibinfo {volume} {11}},\ \bibinfo
  {pages} {361--365} (\bibinfo {year} {2017})}\BibitemShut {NoStop}%
\bibitem [{\citenamefont {Fischer}\ \emph {et~al.}(2016)\citenamefont
  {Fischer}, \citenamefont {M{\"{u}}ller}, \citenamefont {Lagoudakis},\ and\
  \citenamefont {Vuckovic}}]{Fischer2016}%
  \BibitemOpen
  \bibfield  {author} {\bibinfo {author} {\bibfnamefont {K.~A.}\ \bibnamefont
  {Fischer}}, \bibinfo {author} {\bibfnamefont {K.}~\bibnamefont
  {M{\"{u}}ller}}, \bibinfo {author} {\bibfnamefont {K.~G.}\ \bibnamefont
  {Lagoudakis}}, \ and\ \bibinfo {author} {\bibfnamefont {J.}~\bibnamefont
  {Vuckovic}},\ }\bibfield  {title} {\enquote {\bibinfo {title} {{Dynamical
  modeling of pulsed two-photon interference}},}\ }\href@noop {} {\bibfield
  {journal} {\bibinfo  {journal} {New Journal of Physics}\ }\textbf {\bibinfo
  {volume} {18}},\ \bibinfo {pages} {113053} (\bibinfo {year}
  {2016})}\BibitemShut {NoStop}%
\bibitem [{\citenamefont {Fischer}\ \emph
  {et~al.}(2017{\natexlab{a}})\citenamefont {Fischer}, \citenamefont
  {Hanschke}, \citenamefont {Wierzbowski}, \citenamefont {Simmet},
  \citenamefont {Dory}, \citenamefont {Finley}, \citenamefont
  {Vu{\v{c}}kovi{\'{c}}},\ and\ \citenamefont {M{\"{u}}ller}}]{Fischer2017}%
  \BibitemOpen
  \bibfield  {author} {\bibinfo {author} {\bibfnamefont {Kevin~A.}\
  \bibnamefont {Fischer}}, \bibinfo {author} {\bibfnamefont {Lukas}\
  \bibnamefont {Hanschke}}, \bibinfo {author} {\bibfnamefont {Jakob}\
  \bibnamefont {Wierzbowski}}, \bibinfo {author} {\bibfnamefont {Tobias}\
  \bibnamefont {Simmet}}, \bibinfo {author} {\bibfnamefont {Constantin}\
  \bibnamefont {Dory}}, \bibinfo {author} {\bibfnamefont {Jonathan~J.}\
  \bibnamefont {Finley}}, \bibinfo {author} {\bibfnamefont {Jelena}\
  \bibnamefont {Vu{\v{c}}kovi{\'{c}}}}, \ and\ \bibinfo {author} {\bibfnamefont
  {Kai}\ \bibnamefont {M{\"{u}}ller}},\ }\bibfield  {title} {\enquote {\bibinfo
  {title} {{Signatures of two-photon pulses from a quantum two-level
  system}},}\ }\href {\doibase 10.1038/nphys4052} {\bibfield  {journal}
  {\bibinfo  {journal} {Nature Physics}\ }\textbf {\bibinfo {volume} {13}},\
  \bibinfo {pages} {649--654} (\bibinfo {year}
  {2017}{\natexlab{a}})}\BibitemShut {NoStop}%
\bibitem [{\citenamefont {Fischer}\ \emph
  {et~al.}(2017{\natexlab{b}})\citenamefont {Fischer}, \citenamefont
  {Hanschke}, \citenamefont {Kremser}, \citenamefont {Finley}, \citenamefont
  {M{\"u}ller},\ and\ \citenamefont {Vu{\v{c}}kovi{\'c}}}]{Fischer2017-2}%
  \BibitemOpen
  \bibfield  {author} {\bibinfo {author} {\bibfnamefont {Kevin~A}\ \bibnamefont
  {Fischer}}, \bibinfo {author} {\bibfnamefont {Lukas}\ \bibnamefont
  {Hanschke}}, \bibinfo {author} {\bibfnamefont {Malte}\ \bibnamefont
  {Kremser}}, \bibinfo {author} {\bibfnamefont {Jonathan~J}\ \bibnamefont
  {Finley}}, \bibinfo {author} {\bibfnamefont {Kai}\ \bibnamefont
  {M{\"u}ller}}, \ and\ \bibinfo {author} {\bibfnamefont {Jelena}\ \bibnamefont
  {Vu{\v{c}}kovi{\'c}}},\ }\bibfield  {title} {\enquote {\bibinfo {title}
  {Pulsed rabi oscillations in quantum two-level systems: beyond the area
  theorem},}\ }\href@noop {} {\bibfield  {journal} {\bibinfo  {journal}
  {Quantum Science and Technology}\ }\textbf {\bibinfo {volume} {3}},\ \bibinfo
  {pages} {014006} (\bibinfo {year} {2017}{\natexlab{b}})}\BibitemShut
  {NoStop}%
\bibitem [{\citenamefont {Dada}\ \emph {et~al.}(2016)\citenamefont {Dada},
  \citenamefont {Santana}, \citenamefont {Malein}, \citenamefont
  {Koutroumanis}, \citenamefont {Ma}, \citenamefont {Zajac}, \citenamefont
  {Lim}, \citenamefont {Song},\ and\ \citenamefont
  {Gerardot}}]{dada2016indistinguishable}%
  \BibitemOpen
  \bibfield  {author} {\bibinfo {author} {\bibfnamefont {Adetunmise~C}\
  \bibnamefont {Dada}}, \bibinfo {author} {\bibfnamefont {Ted~S}\ \bibnamefont
  {Santana}}, \bibinfo {author} {\bibfnamefont {Ralph~NE}\ \bibnamefont
  {Malein}}, \bibinfo {author} {\bibfnamefont {Antonios}\ \bibnamefont
  {Koutroumanis}}, \bibinfo {author} {\bibfnamefont {Yong}\ \bibnamefont {Ma}},
  \bibinfo {author} {\bibfnamefont {Joanna~M}\ \bibnamefont {Zajac}}, \bibinfo
  {author} {\bibfnamefont {Ju~Y}\ \bibnamefont {Lim}}, \bibinfo {author}
  {\bibfnamefont {Jin~D}\ \bibnamefont {Song}}, \ and\ \bibinfo {author}
  {\bibfnamefont {Brian~D}\ \bibnamefont {Gerardot}},\ }\bibfield  {title}
  {\enquote {\bibinfo {title} {Indistinguishable single photons with flexible
  electronic triggering},}\ }\href@noop {} {\bibfield  {journal} {\bibinfo
  {journal} {Optica}\ }\textbf {\bibinfo {volume} {3}},\ \bibinfo {pages}
  {493--498} (\bibinfo {year} {2016})}\BibitemShut {NoStop}%
\bibitem [{\citenamefont {Bayer}\ \emph {et~al.}(2002)\citenamefont {Bayer},
  \citenamefont {Ortner}, \citenamefont {Stern}, \citenamefont {Kuther},
  \citenamefont {Gorbunov}, \citenamefont {Forchel}, \citenamefont {Hawrylak},
  \citenamefont {Fafard}, \citenamefont {Hinzer}, \citenamefont {Reinecke},
  \citenamefont {Walck}, \citenamefont {Reithmaier}, \citenamefont {Klopf},\
  and\ \citenamefont {Sch{\"{a}}fer}}]{Bayer2002}%
  \BibitemOpen
  \bibfield  {author} {\bibinfo {author} {\bibfnamefont {M.}~\bibnamefont
  {Bayer}}, \bibinfo {author} {\bibfnamefont {G.}~\bibnamefont {Ortner}},
  \bibinfo {author} {\bibfnamefont {O.}~\bibnamefont {Stern}}, \bibinfo
  {author} {\bibfnamefont {A.}~\bibnamefont {Kuther}}, \bibinfo {author}
  {\bibfnamefont {A.~A.}\ \bibnamefont {Gorbunov}}, \bibinfo {author}
  {\bibfnamefont {A.}~\bibnamefont {Forchel}}, \bibinfo {author} {\bibfnamefont
  {P.}~\bibnamefont {Hawrylak}}, \bibinfo {author} {\bibfnamefont
  {S.}~\bibnamefont {Fafard}}, \bibinfo {author} {\bibfnamefont
  {K.}~\bibnamefont {Hinzer}}, \bibinfo {author} {\bibfnamefont {T.~L.}\
  \bibnamefont {Reinecke}}, \bibinfo {author} {\bibfnamefont {S.~N.}\
  \bibnamefont {Walck}}, \bibinfo {author} {\bibfnamefont {J.~P.}\ \bibnamefont
  {Reithmaier}}, \bibinfo {author} {\bibfnamefont {F.}~\bibnamefont {Klopf}}, \
  and\ \bibinfo {author} {\bibfnamefont {F.}~\bibnamefont {Sch{\"{a}}fer}},\
  }\bibfield  {title} {\enquote {\bibinfo {title} {{Fine structure of neutral
  and charged excitons in self-assembled In(Ga)As/(Al)GaAs quantum dots}},}\
  }\href {\doibase 10.1103/PhysRevB.65.195315} {\bibfield  {journal} {\bibinfo
  {journal} {Physical Review B}\ }\textbf {\bibinfo {volume} {65}},\ \bibinfo
  {pages} {195315} (\bibinfo {year} {2002})}\BibitemShut {NoStop}%
\bibitem [{\citenamefont {Finley}\ \emph {et~al.}(2002)\citenamefont {Finley},
  \citenamefont {Mowbray}, \citenamefont {Skolnick}, \citenamefont {Ashmore},
  \citenamefont {Baker}, \citenamefont {Monte},\ and\ \citenamefont
  {Hopkinson}}]{Finley2002}%
  \BibitemOpen
  \bibfield  {author} {\bibinfo {author} {\bibfnamefont {J.~J.}\ \bibnamefont
  {Finley}}, \bibinfo {author} {\bibfnamefont {D.~J.}\ \bibnamefont {Mowbray}},
  \bibinfo {author} {\bibfnamefont {M.~S.}\ \bibnamefont {Skolnick}}, \bibinfo
  {author} {\bibfnamefont {A.~D.}\ \bibnamefont {Ashmore}}, \bibinfo {author}
  {\bibfnamefont {C.}~\bibnamefont {Baker}}, \bibinfo {author} {\bibfnamefont
  {A.~F.~G.}\ \bibnamefont {Monte}}, \ and\ \bibinfo {author} {\bibfnamefont
  {M.}~\bibnamefont {Hopkinson}},\ }\bibfield  {title} {\enquote {\bibinfo
  {title} {{Fine structure of charged and neutral excitons in InAs-AlGaAs
  quantum dots}},}\ }\href {\doibase 10.1103/PhysRevB.66.153316} {\bibfield
  {journal} {\bibinfo  {journal} {Physical Review B}\ }\textbf {\bibinfo
  {volume} {66}},\ \bibinfo {pages} {153316} (\bibinfo {year}
  {2002})}\BibitemShut {NoStop}%
\bibitem [{\citenamefont {Akopian}\ \emph {et~al.}(2006)\citenamefont
  {Akopian}, \citenamefont {Lindner}, \citenamefont {Poem}, \citenamefont
  {Berlatzky}, \citenamefont {Avron}, \citenamefont {Gershoni}, \citenamefont
  {Gerardot},\ and\ \citenamefont {Petroff}}]{Akopian2006}%
  \BibitemOpen
  \bibfield  {author} {\bibinfo {author} {\bibfnamefont {N.}~\bibnamefont
  {Akopian}}, \bibinfo {author} {\bibfnamefont {N.~H.}\ \bibnamefont
  {Lindner}}, \bibinfo {author} {\bibfnamefont {E.}~\bibnamefont {Poem}},
  \bibinfo {author} {\bibfnamefont {Y.}~\bibnamefont {Berlatzky}}, \bibinfo
  {author} {\bibfnamefont {J.}~\bibnamefont {Avron}}, \bibinfo {author}
  {\bibfnamefont {D.}~\bibnamefont {Gershoni}}, \bibinfo {author}
  {\bibfnamefont {B.~D.}\ \bibnamefont {Gerardot}}, \ and\ \bibinfo {author}
  {\bibfnamefont {P.~M.}\ \bibnamefont {Petroff}},\ }\bibfield  {title}
  {\enquote {\bibinfo {title} {{Entangled photon pairs from semiconductor
  quantum dots}},}\ }\href {\doibase 10.1103/PhysRevLett.96.130501} {\bibfield
  {journal} {\bibinfo  {journal} {Physical Review Letters}\ }\textbf {\bibinfo
  {volume} {96}},\ \bibinfo {pages} {130501} (\bibinfo {year}
  {2006})}\BibitemShut {NoStop}%
\bibitem [{\citenamefont {M{\"{u}}ller}\ \emph {et~al.}(2014)\citenamefont
  {M{\"{u}}ller}, \citenamefont {Bounouar}, \citenamefont {J{\"{o}}ns},
  \citenamefont {Gl{\"{a}}ssl},\ and\ \citenamefont {Michler}}]{Mueller2014}%
  \BibitemOpen
  \bibfield  {author} {\bibinfo {author} {\bibfnamefont {M.}~\bibnamefont
  {M{\"{u}}ller}}, \bibinfo {author} {\bibfnamefont {S.}~\bibnamefont
  {Bounouar}}, \bibinfo {author} {\bibfnamefont {K.~D.}\ \bibnamefont
  {J{\"{o}}ns}}, \bibinfo {author} {\bibfnamefont {M.}~\bibnamefont
  {Gl{\"{a}}ssl}}, \ and\ \bibinfo {author} {\bibfnamefont {P.}~\bibnamefont
  {Michler}},\ }\bibfield  {title} {\enquote {\bibinfo {title} {{On-demand
  generation of indistinguishable polarization-entangled photon pairs}},}\
  }\href {\doibase 10.1038/nphoton.2013.377} {\bibfield  {journal} {\bibinfo
  {journal} {Nature Photonics}\ }\textbf {\bibinfo {volume} {8}},\ \bibinfo
  {pages} {224--228} (\bibinfo {year} {2014})}\BibitemShut {NoStop}%
\bibitem [{\citenamefont {Huber}\ \emph {et~al.}(2017)\citenamefont {Huber},
  \citenamefont {Reindl}, \citenamefont {Huo}, \citenamefont {Huang},
  \citenamefont {Wildmann}, \citenamefont {Schmidt}, \citenamefont {Rastelli},\
  and\ \citenamefont {Trotta}}]{Huber2017}%
  \BibitemOpen
  \bibfield  {author} {\bibinfo {author} {\bibfnamefont {Daniel}\ \bibnamefont
  {Huber}}, \bibinfo {author} {\bibfnamefont {Marcus}\ \bibnamefont {Reindl}},
  \bibinfo {author} {\bibfnamefont {Yongheng}\ \bibnamefont {Huo}}, \bibinfo
  {author} {\bibfnamefont {Huiying}\ \bibnamefont {Huang}}, \bibinfo {author}
  {\bibfnamefont {Johannes~S.}\ \bibnamefont {Wildmann}}, \bibinfo {author}
  {\bibfnamefont {Oliver~G.}\ \bibnamefont {Schmidt}}, \bibinfo {author}
  {\bibfnamefont {Armando}\ \bibnamefont {Rastelli}}, \ and\ \bibinfo {author}
  {\bibfnamefont {Rinaldo}\ \bibnamefont {Trotta}},\ }\bibfield  {title}
  {\enquote {\bibinfo {title} {{Highly indistinguishable and strongly entangled
  photons from symmetric GaAs quantum dots}},}\ }\href {\doibase
  10.1038/ncomms15506} {\bibfield  {journal} {\bibinfo  {journal} {Nature
  Communications}\ }\textbf {\bibinfo {volume} {8}},\ \bibinfo {pages} {15506}
  (\bibinfo {year} {2017})}\BibitemShut {NoStop}%
\bibitem [{\citenamefont {Ardelt}\ \emph {et~al.}(2014)\citenamefont {Ardelt},
  \citenamefont {Hanschke}, \citenamefont {Fischer}, \citenamefont
  {M{\"{u}}ller}, \citenamefont {Kleinkauf}, \citenamefont {Koller},
  \citenamefont {Bechtold}, \citenamefont {Simmet}, \citenamefont
  {Wierzbowski}, \citenamefont {Riedl}, \citenamefont {Abstreiter},\ and\
  \citenamefont {Finley}}]{Ardelt2014}%
  \BibitemOpen
  \bibfield  {author} {\bibinfo {author} {\bibfnamefont {Per~Lennart}\
  \bibnamefont {Ardelt}}, \bibinfo {author} {\bibfnamefont {Lukas}\
  \bibnamefont {Hanschke}}, \bibinfo {author} {\bibfnamefont {Kevin~A.}\
  \bibnamefont {Fischer}}, \bibinfo {author} {\bibfnamefont {Kai}\ \bibnamefont
  {M{\"{u}}ller}}, \bibinfo {author} {\bibfnamefont {Alexander}\ \bibnamefont
  {Kleinkauf}}, \bibinfo {author} {\bibfnamefont {Manuel}\ \bibnamefont
  {Koller}}, \bibinfo {author} {\bibfnamefont {Alexander}\ \bibnamefont
  {Bechtold}}, \bibinfo {author} {\bibfnamefont {Tobias}\ \bibnamefont
  {Simmet}}, \bibinfo {author} {\bibfnamefont {Jakob}\ \bibnamefont
  {Wierzbowski}}, \bibinfo {author} {\bibfnamefont {Hubert}\ \bibnamefont
  {Riedl}}, \bibinfo {author} {\bibfnamefont {Gerhard}\ \bibnamefont
  {Abstreiter}}, \ and\ \bibinfo {author} {\bibfnamefont {Jonathan~J.}\
  \bibnamefont {Finley}},\ }\bibfield  {title} {\enquote {\bibinfo {title}
  {{Dissipative preparation of the exciton and biexciton in self-assembled
  quantum dots on picosecond time scales}},}\ }\href {\doibase
  10.1103/PhysRevB.90.241404} {\bibfield  {journal} {\bibinfo  {journal}
  {Physical Review B}\ }\textbf {\bibinfo {volume} {90}},\ \bibinfo {pages}
  {241404} (\bibinfo {year} {2014})}\BibitemShut {NoStop}%
\bibitem [{\citenamefont {Schweickert}\ \emph {et~al.}(2018)\citenamefont
  {Schweickert}, \citenamefont {J\"{o}ns}, \citenamefont {Zeuner},
  \citenamefont {Covre~da Silva}, \citenamefont {Huang}, \citenamefont
  {Lettner}, \citenamefont {Reindl}, \citenamefont {Zichi}, \citenamefont
  {Trotta}, \citenamefont {Rastelli},\ and\ \citenamefont
  {Zwiller}}]{Schweickert2017}%
  \BibitemOpen
  \bibfield  {author} {\bibinfo {author} {\bibfnamefont {Lucas}\ \bibnamefont
  {Schweickert}}, \bibinfo {author} {\bibfnamefont {Klaus~D.}\ \bibnamefont
  {J\"{o}ns}}, \bibinfo {author} {\bibfnamefont {Katharina~D.}\ \bibnamefont
  {Zeuner}}, \bibinfo {author} {\bibfnamefont {Saimon~Filipe}\ \bibnamefont
  {Covre~da Silva}}, \bibinfo {author} {\bibfnamefont {Huiying}\ \bibnamefont
  {Huang}}, \bibinfo {author} {\bibfnamefont {Thomas}\ \bibnamefont {Lettner}},
  \bibinfo {author} {\bibfnamefont {Marcus}\ \bibnamefont {Reindl}}, \bibinfo
  {author} {\bibfnamefont {Julien}\ \bibnamefont {Zichi}}, \bibinfo {author}
  {\bibfnamefont {Rinaldo}\ \bibnamefont {Trotta}}, \bibinfo {author}
  {\bibfnamefont {Armando}\ \bibnamefont {Rastelli}}, \ and\ \bibinfo {author}
  {\bibfnamefont {Val}\ \bibnamefont {Zwiller}},\ }\bibfield  {title} {\enquote
  {\bibinfo {title} {On-demand generation of background-free single photons
  from a solid-state source},}\ }\href {\doibase 10.1063/1.5020038} {\bibfield
  {journal} {\bibinfo  {journal} {Applied Physics Letters}\ }\textbf {\bibinfo
  {volume} {112}},\ \bibinfo {pages} {093106} (\bibinfo {year}
  {2018})}\BibitemShut {NoStop}%
\bibitem [{\citenamefont {Shore}(2011)}]{Shore2011-fz}%
  \BibitemOpen
  \bibfield  {author} {\bibinfo {author} {\bibfnamefont {Bruce~W.}\
  \bibnamefont {Shore}},\ }\href@noop {} {\emph {\bibinfo {title} {Manipulating
  quantum structures using laser pulses}}}\ (\bibinfo  {publisher} {Cambridge
  University Press, Cambridge},\ \bibinfo {year} {2011})\BibitemShut {NoStop}%
\bibitem [{\citenamefont {Johansson}\ \emph {et~al.}(2013)\citenamefont
  {Johansson}, \citenamefont {Nation},\ and\ \citenamefont
  {Nori}}]{Johansson2014}%
  \BibitemOpen
  \bibfield  {author} {\bibinfo {author} {\bibfnamefont {J.R.}\ \bibnamefont
  {Johansson}}, \bibinfo {author} {\bibfnamefont {P.D.}\ \bibnamefont
  {Nation}}, \ and\ \bibinfo {author} {\bibfnamefont {Franco}\ \bibnamefont
  {Nori}},\ }\bibfield  {title} {\enquote {\bibinfo {title} {Qutip 2: A python
  framework for the dynamics of open quantum systems},}\ }\href {\doibase
  10.1016/j.cpc.2012.11.019} {\bibfield  {journal} {\bibinfo  {journal}
  {Computer Physics Communications}\ }\textbf {\bibinfo {volume} {184}},\
  \bibinfo {pages} {1234 -- 1240} (\bibinfo {year} {2013})}\BibitemShut
  {NoStop}%
\bibitem [{\citenamefont {Fischer}\ \emph {et~al.}(2018)\citenamefont
  {Fischer}, \citenamefont {Trivedi}, \citenamefont {Ramasesh}, \citenamefont
  {Siddiqi},\ and\ \citenamefont {Vu{\v{c}}kovi{\'c}}}]{fischer2017scattering}%
  \BibitemOpen
  \bibfield  {author} {\bibinfo {author} {\bibfnamefont {Kevin~A}\ \bibnamefont
  {Fischer}}, \bibinfo {author} {\bibfnamefont {Rahul}\ \bibnamefont
  {Trivedi}}, \bibinfo {author} {\bibfnamefont {Vinay}\ \bibnamefont
  {Ramasesh}}, \bibinfo {author} {\bibfnamefont {Irfan}\ \bibnamefont
  {Siddiqi}}, \ and\ \bibinfo {author} {\bibfnamefont {Jelena}\ \bibnamefont
  {Vu{\v{c}}kovi{\'c}}},\ }\bibfield  {title} {\enquote {\bibinfo {title}
  {Scattering into one-dimensional waveguides from a coherently-driven
  quantum-optical system},}\ }\href@noop {} {\bibfield  {journal} {\bibinfo
  {journal} {Quantum}\ }\textbf {\bibinfo {volume} {2}},\ \bibinfo {pages} {69}
  (\bibinfo {year} {2018})}\BibitemShut {NoStop}%
\bibitem [{\citenamefont {Carmichael}(2009)}]{carmichael2009open}%
  \BibitemOpen
  \bibfield  {author} {\bibinfo {author} {\bibfnamefont {Howard}\ \bibnamefont
  {Carmichael}},\ }\href@noop {} {\emph {\bibinfo {title} {An open systems
  approach to quantum optics: lectures presented at the Universit{\'e} Libre de
  Bruxelles, October 28 to November 4, 1991}}},\ Vol.~\bibinfo {volume} {18}\
  (\bibinfo  {publisher} {Springer Science \& Business Media},\ \bibinfo {year}
  {2009})\BibitemShut {NoStop}%
\bibitem [{\citenamefont {Wiseman}\ and\ \citenamefont
  {Milburn}(2009)}]{wiseman2009quantum}%
  \BibitemOpen
  \bibfield  {author} {\bibinfo {author} {\bibfnamefont {Howard~M}\
  \bibnamefont {Wiseman}}\ and\ \bibinfo {author} {\bibfnamefont {Gerard~J}\
  \bibnamefont {Milburn}},\ }\href@noop {} {\emph {\bibinfo {title} {Quantum
  measurement and control}}}\ (\bibinfo  {publisher} {Cambridge university
  press},\ \bibinfo {year} {2009})\BibitemShut {NoStop}%
\bibitem [{\citenamefont {Gardiner}\ and\ \citenamefont
  {Zoller}(2015)}]{gardiner2015quantum}%
  \BibitemOpen
  \bibfield  {author} {\bibinfo {author} {\bibfnamefont {Crispin}\ \bibnamefont
  {Gardiner}}\ and\ \bibinfo {author} {\bibfnamefont {Peter}\ \bibnamefont
  {Zoller}},\ }\bibfield  {title} {\enquote {\bibinfo {title} {The quantum
  world of ultra-cold atoms and light book ii: The physics of quantum-optical
  devices},}\ }in\ \href@noop {} {\emph {\bibinfo {booktitle} {The Quantum
  World of Ultra-Cold Atoms and Light Book II: The Physics of Quantum-Optical
  Devices}}}\ (\bibinfo  {publisher} {World Scientific},\ \bibinfo {year}
  {2015})\ pp.\ \bibinfo {pages} {1--524}\BibitemShut {NoStop}%
\bibitem [{\citenamefont {Gazzano}\ \emph {et~al.}(2013)\citenamefont
  {Gazzano}, \citenamefont {De~Vasconcellos}, \citenamefont {Arnold},
  \citenamefont {Nowak}, \citenamefont {Galopin}, \citenamefont {Sagnes},
  \citenamefont {Lanco}, \citenamefont {Lema{\^\i}tre},\ and\ \citenamefont
  {Senellart}}]{Gazzano2013}%
  \BibitemOpen
  \bibfield  {author} {\bibinfo {author} {\bibfnamefont {O}~\bibnamefont
  {Gazzano}}, \bibinfo {author} {\bibfnamefont {S~Michaelis}\ \bibnamefont
  {De~Vasconcellos}}, \bibinfo {author} {\bibfnamefont {C}~\bibnamefont
  {Arnold}}, \bibinfo {author} {\bibfnamefont {A}~\bibnamefont {Nowak}},
  \bibinfo {author} {\bibfnamefont {E}~\bibnamefont {Galopin}}, \bibinfo
  {author} {\bibfnamefont {I}~\bibnamefont {Sagnes}}, \bibinfo {author}
  {\bibfnamefont {L}~\bibnamefont {Lanco}}, \bibinfo {author} {\bibfnamefont
  {A}~\bibnamefont {Lema{\^\i}tre}}, \ and\ \bibinfo {author} {\bibfnamefont
  {P}~\bibnamefont {Senellart}},\ }\bibfield  {title} {\enquote {\bibinfo
  {title} {Bright solid-state sources of indistinguishable single photons},}\
  }\href@noop {} {\bibfield  {journal} {\bibinfo  {journal} {Nature
  communications}\ }\textbf {\bibinfo {volume} {4}},\ \bibinfo {pages} {1425}
  (\bibinfo {year} {2013})}\BibitemShut {NoStop}%
\bibitem [{\citenamefont {Claudon}\ \emph {et~al.}(2010)\citenamefont
  {Claudon}, \citenamefont {Bleuse}, \citenamefont {Malik}, \citenamefont
  {Bazin}, \citenamefont {Jaffrennou}, \citenamefont {Gregersen}, \citenamefont
  {Sauvan}, \citenamefont {Lalanne},\ and\ \citenamefont
  {G\'{e}rard}}]{Claudon2010}%
  \BibitemOpen
  \bibfield  {author} {\bibinfo {author} {\bibfnamefont {Julien}\ \bibnamefont
  {Claudon}}, \bibinfo {author} {\bibfnamefont {Jo\"{e}l}\ \bibnamefont
  {Bleuse}}, \bibinfo {author} {\bibfnamefont {Nitin~S.}\ \bibnamefont
  {Malik}}, \bibinfo {author} {\bibfnamefont {Maela}\ \bibnamefont {Bazin}},
  \bibinfo {author} {\bibfnamefont {P\'{e}rine}\ \bibnamefont {Jaffrennou}},
  \bibinfo {author} {\bibfnamefont {Niels}\ \bibnamefont {Gregersen}}, \bibinfo
  {author} {\bibfnamefont {Christophe}\ \bibnamefont {Sauvan}}, \bibinfo
  {author} {\bibfnamefont {Philippe}\ \bibnamefont {Lalanne}}, \ and\ \bibinfo
  {author} {\bibfnamefont {Jean-Michel}\ \bibnamefont {G\'{e}rard}},\
  }\bibfield  {title} {\enquote {\bibinfo {title} {{A highly efficient
  single-photon source based on a quantum dot in a photonic nanowire}},}\
  }\href {\doibase 10.1038/nphoton.2009.287} {\bibfield  {journal} {\bibinfo
  {journal} {Nature Photonics}\ }\textbf {\bibinfo {volume} {4}},\ \bibinfo
  {pages} {174--177} (\bibinfo {year} {2010})}\BibitemShut {NoStop}%
\bibitem [{\citenamefont {Gschrey}\ \emph {et~al.}(2015)\citenamefont
  {Gschrey}, \citenamefont {Thoma}, \citenamefont {Schnauber}, \citenamefont
  {Seifried}, \citenamefont {Schmidt}, \citenamefont {Wohlfeil}, \citenamefont
  {Kruger}, \citenamefont {Schulze}, \citenamefont {Heindel}, \citenamefont
  {Burger}, \citenamefont {Schmidt}, \citenamefont {Strittmatter},
  \citenamefont {Rodt},\ and\ \citenamefont {Reitzenstein}}]{Gschrey2015}%
  \BibitemOpen
  \bibfield  {author} {\bibinfo {author} {\bibfnamefont {M.}~\bibnamefont
  {Gschrey}}, \bibinfo {author} {\bibfnamefont {A.}~\bibnamefont {Thoma}},
  \bibinfo {author} {\bibfnamefont {P.}~\bibnamefont {Schnauber}}, \bibinfo
  {author} {\bibfnamefont {M.}~\bibnamefont {Seifried}}, \bibinfo {author}
  {\bibfnamefont {R.}~\bibnamefont {Schmidt}}, \bibinfo {author} {\bibfnamefont
  {B.}~\bibnamefont {Wohlfeil}}, \bibinfo {author} {\bibfnamefont
  {L.}~\bibnamefont {Kruger}}, \bibinfo {author} {\bibfnamefont {J.~H.}\
  \bibnamefont {Schulze}}, \bibinfo {author} {\bibfnamefont {T.}~\bibnamefont
  {Heindel}}, \bibinfo {author} {\bibfnamefont {S.}~\bibnamefont {Burger}},
  \bibinfo {author} {\bibfnamefont {F.}~\bibnamefont {Schmidt}}, \bibinfo
  {author} {\bibfnamefont {A.}~\bibnamefont {Strittmatter}}, \bibinfo {author}
  {\bibfnamefont {S.}~\bibnamefont {Rodt}}, \ and\ \bibinfo {author}
  {\bibfnamefont {S.}~\bibnamefont {Reitzenstein}},\ }\bibfield  {title}
  {\enquote {\bibinfo {title} {Highly indistinguishable photons from
  deterministic quantum-dot microlenses utilizing three-dimensional in situ
  electron-beam lithography},}\ }\href {http://dx.doi.org/10.1038/ncomms8662}
  {\bibfield  {journal} {\bibinfo  {journal} {Nature Communications}\ }\textbf
  {\bibinfo {volume} {6}} (\bibinfo {year} {2015})}\BibitemShut {NoStop}%
\bibitem [{\citenamefont {Sapienza}\ \emph {et~al.}(2015)\citenamefont
  {Sapienza}, \citenamefont {Davanco}, \citenamefont {Badolato},\ and\
  \citenamefont {Srinivasan}}]{Sapienza2015}%
  \BibitemOpen
  \bibfield  {author} {\bibinfo {author} {\bibfnamefont {Luca}\ \bibnamefont
  {Sapienza}}, \bibinfo {author} {\bibfnamefont {Marcelo}\ \bibnamefont
  {Davanco}}, \bibinfo {author} {\bibfnamefont {Antonio}\ \bibnamefont
  {Badolato}}, \ and\ \bibinfo {author} {\bibfnamefont {Kartik}\ \bibnamefont
  {Srinivasan}},\ }\bibfield  {title} {\enquote {\bibinfo {title} {Dots for
  bright and pure single-photon emission},}\ }\href {\doibase
  10.1038/ncomms8833} {\bibfield  {journal} {\bibinfo  {journal} {Nature
  Communications}\ }\textbf {\bibinfo {volume} {6}},\ \bibinfo {pages} {7833}
  (\bibinfo {year} {2015})}\BibitemShut {NoStop}%
\end{thebibliography}%

\end{document}